\documentclass[aip, amsmath,amssymb, preprint]{revtex4-1}

\usepackage{CJK}
\usepackage[final]{graphicx}
\usepackage{hyperref}
\usepackage{color}
\usepackage{xr}
\usepackage{dcolumn}
\usepackage{bm}
\usepackage[utf8]{inputenc}
\usepackage[T1]{fontenc}
\usepackage{mathptmx}
\usepackage{etoolbox}
\usepackage{hyperref}
\usepackage{microtype}

\makeatletter
\def\@email#1#2{%
 \endgroup
 \patchcmd{\titleblock@produce}
  {\frontmatter@RRAPformat}
  {\frontmatter@RRAPformat{\produce@RRAP{*#1\href{mailto:#2}{#2}}}\frontmatter@RRAPformat}
  {}{}
}%
\makeatother

\begin{document}
\begin{CJK*}{GB}{} 

\preprint{STS-Pionner/Incident-Beam-Optics-Optimization}

\title[]{Incident beam optics optimization for the single crystal neutron diffractometer Pioneer with a polarized beam option}
\author{Yaohua Liu*}
\email[Author to whom correspondence should be addressed: ]{liuyh@ornl.gov}
\affiliation{Second Target Station, Oak Ridge National Laboratory, Oak Ridge, Tennessee 37831, USA}%
\author{Peter Torres}
\affiliation{Second Target Station, Oak Ridge National Laboratory, Oak Ridge, Tennessee 37831, USA}%

\date{\today}
\begin{abstract}
Pioneer, a next-generation single-crystal neutron diffractometer, is under development for Oak Ridge National Laboratory's Second Target Station (STS). Designed to address a wide range of scientific questions, Pioneer will deliver homogeneous neutron beams with customizable size and divergence, and provide a polarized beam option. This article introduces its incident beam optics, highlighting the optimization methodology and the simulated performance. Pioneer will utilize a modified elliptical-straight guide for neutron transport and deploy slit packages and insertable apertures to control beam size and divergence. The optimized guide geometry matches the optimal-and-full-sample-illumination condition, and the beam control system effectively filters out unwanted neutrons while preserving the desired ones. Additionally, we have found that polygon-approximated guides provide satisfactory transport efficiency and beam homogeneity, eliminating the need of truly curved guides. To enhance neutronics performance and reduce cost, the coatings of supermirror elements are individually optimized to the lowest half-integer $m$-values that are sufficient to deliver the desired neutrons. After evaluating polarizing V-cavities and $^3$He spin filters over the default polarized wavelength band of 1.2-5.5~\AA, we selected a translatable multichannel polarizing V-cavity as the incident beam polarizer. Strategically placed at a location where the beam divergence is low and a large in-guide gap has negligible impact on transport efficiency, the optimized V-cavity achieves an average $P^2T$ of approximately 35\%. 

\end{abstract}
\maketitle
\end{CJK*}

\section{introduction}
Neutron diffraction is a cornerstone for unraveling the intricate structure-property relationships of advanced materials. There are enduring needs to probe smaller-volume samples, detect weaker signals, and utilize versatile sample environments. The Second Target Station (STS) at Oak Ridge National Laboratory will use novel, compact moderators, combined with short proton pulses striking the target, to produce world-leading peak brightness cold neutron pulses at 15~Hz~\cite{adams2020first}. Pioneer, a next-generation single-crystal neutron diffractometer, aims to addresses these challenges by leveraging the high brilliance source at STS and incorporating modern instrument design. Pioneer will provide a tenfold or greater increase in neutron flux on the sample than existing TOF single-crystal neutron diffractometers at comparable resolutions, such as TOPAZ at the First Target Station and SENJU at J-PARC.  Employing a time-of-flight technique with a wavelength bandwidth of 4.3~\AA~and a detector coverage of 6.6 steradians, Pioneer will simultaneously sample a substantial reciprocal space volume, facilitating the study of both Bragg diffraction and diffuse scattering from crystalline materials~\cite{wilson2000single, nield2001diffuse, carpenter2015elements}. With its exceptional sensitivity to weak signals, Pioneer will be capable of measuring X-ray-sized single crystals as small as 0.001~mm$^3$ and ultra-thin films of 10~nm thickness. Moreover, Pioneer's ability to collect high-quality data under versatile sample environments positions it as a valuable tool for accelerating materials discovery~\cite{liu2022Pioneer}.

Pioneer requires homogeneous beams with configurable size and divergence and a polarized incident beam option to match various experimental needs. To achieve this, we have conducted Monte Carlo ray tracing simulations to optimize the incident beam optics. The instrument will utilize a modified elliptical-straight guide for neutron transport. Slit packages and insertable apertures will be used to manipulate the phase space at the sample position, providing an operational friendly approach to choose between high-flux modes for studying small samples in versatile sample environments and high-resolution modes for measuring complex crystalline materials with long characteristic length scales. The guide system contains multiple large in-guide gaps to accommodate a T$_0$ chopper and three oversized steel collimators to block high-energy neutrons and gamma rays~\cite{carpenter2015elements}.

To optimize beam transport and control, we selected three representative scenarios with distinct beam size and divergence requirements. Post-optimization analysis confirms that the optimized guide geometry satisfies the optimal-and-full-sample illumination condition~\cite{konik2023new}, and the beam control system effectively suppresses unwanted neutrons. To minimize costs and downstream shielding requirements, we selected the minimum m-values for individual supermirror elements required to transport the desired neutrons.The entire neutron transport and control system, with the exception of a few small gaps, is maintained under vacuum.

Pioneer will utilize a translatable multichannel V-cavity to polarize the incident beam across the default polarized band of 1.2-5.5~\AA. The location and configuration of the V-cavity have been carefully optimized, which shows superior performance compared to $^3$He neutron spin filters over this broad band. Provisions have also been made for a future upgrade to incorporate $^3$He neutron spin filters, extending polarized beam capabilities down to 0.8~\AA~for specific applications, such as accurately determining the magnetic form factor in materials with strong covalency effects~\cite{walters2009effect}.

This paper presents the optimization methodology and simulated performance of Pioneer's incident beam optics. The chopper systems, including their types and locations, and the rationale behind various design choices for Pioneer are reported elsewhere~\cite{liu2025optical}. Section~\ref{sec:spec} introduces the science requirements that drive the design. Section~\ref{sec:guide} focuses on the guide geometry, the beam size and divergence control system, and the guide coating. Section~\ref{sec:pol} discusses the incident-beam polarization system. Finally, Sec.~\ref{sec:discussion} presents key lessons learned and summarizes the findings.

\section{Instrument specification}{\label{sec:spec}}
Pioneer requires a minimum operational wavelength range of 1.0 to 6.0~\AA. During operations, the chopper system will select a 4.3~\AA~band. The instrument design for Pioneer has undergone significant changes from its original concept~\cite{liu2022Pioneer}. The guide system has been transitioned from Montel mirrors to a straight-elliptical guide configuration to enhance thermal neutron transport, following a comparative study on several guide options~\cite{liu2024general}. The source-to-sample distance ($L_1$) is reduced from 60.0~m to 56.0~m, increasing the wavelength bandwidth and avoiding interference with the STS building. Additionally, the maximum beam divergence is increased to enhance flux on samples. Pioneer will use magnetic-field-insensitive silicon photomultiplier (SiPM) cameras for neutron detection. To achieve a high data collection efficiency, Pioneer will install a large vertical cylindrical detector array (5.9 sr) and a removable, flat detector array (0.7 sr) beneath the sample position~\cite{liu2025optical}.

To support diverse scientific missions, Pioneer requires homogeneous beam profiles at the sample position with customizable sizes ranging from 1 to 5~mm and tunable divergences up to 0.7$^\circ$ (full-width-at-half-maximum, FWHM). The beam size is defined by the region with a uniform flux profile, outside of which the flux shall be minimized to reduce background. While Pioneer will frequently study sub-mm crystals, larger samples will be used for thin-film diffraction, time-resolved studies, and pulsed-field experiments. Large beam divergences will enable high-flux (HF) modes to investigate small-volume samples and weak signals, while small beam divergences will support high-resolution (HR) modes to probe complex materials with long characteristic length scales.

Furthermore, Pioneer requires a polarized incident beam option for local magnetic susceptibility studies. The default minimum wavelength $\lambda_{min}$ is 1.2~\AA~ for the polarized beam, with a high figure-of-metric (FOM) $P^2T$ of $\sim 35\%$, averaged over the full band of 1.2-5.5~\AA, where $P$ represents beam polarization and $T$ is the transmission of the polarizer.

\begin{figure*}[t]
\includegraphics[width=0.96\textwidth]{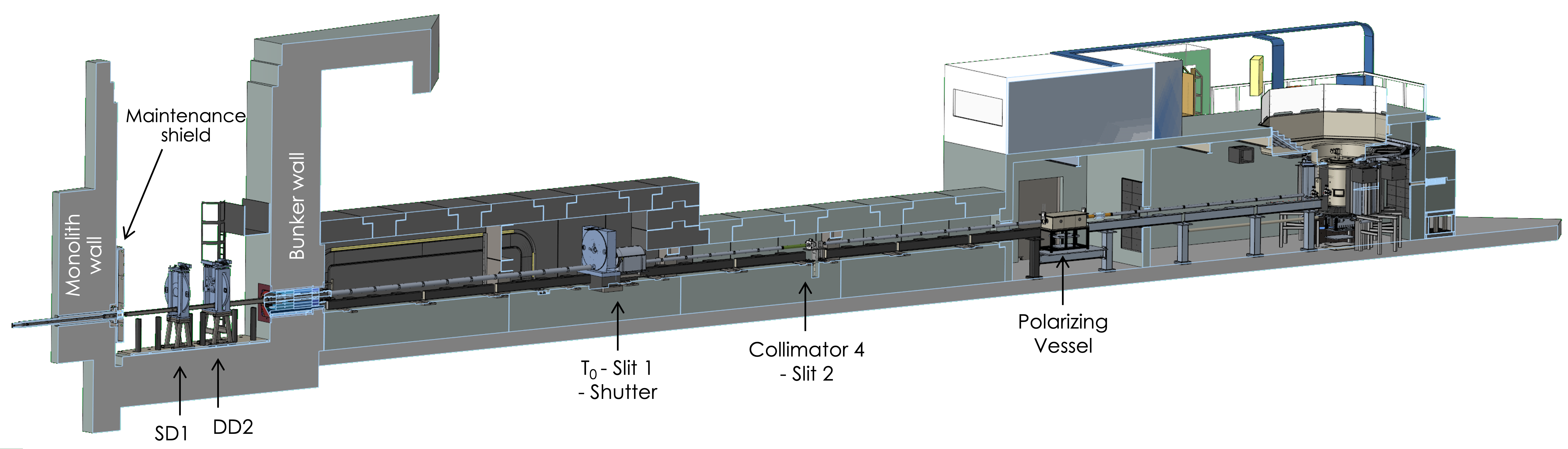}
\caption{\label{fig:layout} Sectional view of the Pioneer beamline. The neutron transport system design shall consider the constraints from the source, the optics, and the building. The monolith and bunker walls create breaks in the guide system, while multiple optical components introduce significant in-guide gaps. See the main text and Tab.~\ref{tab:gaps} for details.}
\end{figure*}

\begin{table*}[ht!]
\caption{\label{tab:gaps} In-guide gap sizes and associated optics. Only components or assemblies causing gaps larger than 0.10~m are included. Optical positions are specified using either start and end coordinates or a single central coordinate. The T$_0$ chopper is 0.45-m long, including a 0.30-m Inconel block. Each slit is 0.23-m long. Of the four collimators, C2 is 0.22-m long, and C3 through C5 are each 1.00-m long. }
{\small
\centering
\begin{tabular}{>{\raggedright\arraybackslash}p{4.6cm}|>{\centering\arraybackslash}p{2.8cm}|>{\centering\arraybackslash}p{2.8cm}|>{\raggedright\arraybackslash}p{5.5cm}}
\hline \hline
\hspace{48pt}\textbf{Name}          & \textbf{Gap Size (m)} & \textbf{Position (m)}     & \hspace{58pt}\textbf{Comment}                      \\ \hline
Maintenance Shield Assembly         & 0.33                 & 5.25--5.58                 &                                           \\ 
\hspace{10pt} Maintenance Shield    &                      & 5.40                 & Interchangeable with Collimator 2         \\ \hline
Disk Chopper 1 Assembly             & 0.24                 & 7.58--7.82                 &                                            \\ 
\hspace{10pt} Monitor 1             &                      & 7.71                      & Before SD1; retractable                     \\ 
\hspace{10pt} Single-disk Chopper   &                      & 7.78                      & SD1                                         \\ \hline
Disk Chopper 2 Assembly             & 0.27                 & 9.32--9.59                 &                                        \\ 
\hspace{10pt} Double-disk Chopper   &                      & 9.38                      & DD2                                      \\ 
\hspace{10pt} Monitor 2             &                      & 9.56                      & After DD2; retractable                 \\ \hline
T$_0$-Shutter Assembly              &    1.91                 & 24.92--26.83               &                    \\ 
\hspace{10pt} T$_0$~Chopper         &                      & 25.10                     &                                         \\ 
\hspace{10pt} Slit 1                &                      & 25.70                     &                                        \\
\hspace{10pt} shutter     &                      & 26.31                     & Interchangeable with Collimator 3         \\ \hline
Collimator 4 - Slit 2 Assembly      & 1.74                 & 32.83-34.57               & Soft constraint                                      \\ 
\hspace{10pt}Collimator 4           &                      & 33.33                     &                                        \\
\hspace{10pt}Slit 2                 &                      & 34.00                     & Retractable                           \\ 
\hspace{10pt}Monitor 3              &                      & 34.20                     &                                        \\ \hline
Polarizing Vessel Assembly          & 1.89                 & 43.57--45.46              &                                        \\ 
\hspace{10pt} Pre-vessel Gate Valve &                      & 43.68                     &                                        \\ 
\hspace{10pt} Monitor 4             &                      & 43.90                     & Retractable                           \\
\hspace{10pt} V-Cavity              &                      & 44.60                     & Interchangeable with Collimator 5         \\ 
\hspace{10pt} Post-vessel Gate Valve&                      & 45.35                     &                                        \\ \hline
\end{tabular}
}
\end{table*}

\section{Neutron Transport and Control System}{\label{sec:guide}}
This section focuses on the optics used to transport neutrons from the moderator to the sample position within the desired phase space. Optics specific to the polarized beam will be addressed later in Sec.\ref{sec:pol}. Neutron guides, which consist of neutron supermirrors with highly reflective surfaces, are used for long-distance beam transport with minimal loss. Since Ferenc Mezei introduced the concept of neutron supermirrors\cite{mezei1976novel}, improvements in manufacturing technology have increased the maximum critical wavevector transfer and reduced reflection losses. Guide geometries have also evolved beyond the original constant rectangular cross-section designs. Modern commercially available options include linearly tapered guides, truly curved guides, and guides with octagonal cross-sections. Simultaneously, advances in Monte Carlo ray tracing software, such as McStas~\cite{willendrup2014mcstas}, VITESS~\cite{lieutenant2004neutron}, and MCViNE~\cite{lin2016MCViNE}, enable high-fidelity simulations and efficient optimization of guides with complex geometries while accommodating various engineering constraints.

Section~\ref{subsec:constraints} introduces the constraints guiding the design of Pioneer's incident beam optics. The methodologies for optimizing the guide geometry and coating are described in Secs.~\ref{subsec:guide} and~\ref{subsec:supermirror}, respectively. Section~\ref{subsec:guideopt} describes the optimized guide geometry. The transport and control performance after optimization is presented in Sec.~\ref{subsec:transport}.

\subsection{Engineering Reality}{\label{subsec:constraints}}
Figure~\ref{fig:layout} shows a sectional view of Pioneer's instrument. The optical design for the neutron transport system must adhere to the engineering constraints imposed by the source, the optics, and the building. For Pioneer, the moderator-to-sample distance is fixed to be 56.0~m, balancing the required wavelength resolution, wavelength bandwidth and STS building limitations. Neutrons are transported from the moderator to the sample position via neutron guides. Along the path, the monolith and bunker walls introduce breaks in the guide system to accommodate inserts that house the guide to pass through the walls. Furthermore, substantial spaces are required to accommodate the maintenance shield (also known as the primary shutter), choppers, collimators, polarizer, slits, monitors, and gate valves. To control the neutron beam, Pioneer uses four slit packages and a pre-sample aperture to manipulate the phase space. Additionally, a polarizing V-cavity is utilized to polarize the incident beam.

As discussed previously~\cite{liu2025optical}, we opt not to use translatable guides for Pioneer because the small guide cross-section imposes tight alignment requirements. Instead, oversized in-guide collimators will be translated into the beam path to block unwanted neutrons when components such as the maintenance shield, shutter, and polarizing V-cavity are not in use. This design results in multiple significant in-guide gaps. Without a carefully designed guide, these large gaps could lead to significant flux losses and beam profile inhomogeneity. Table~\ref{tab:gaps} lists major in-guide gaps and associated optical components. Slits 3 and 4 are not included in the table since they are located after the guide and do not constrain the guide design.  The gap related to the Collimator 4 - Slit 2 assembly is not set by the required space from the optics, rather it is determined by the guide optimization and our strategy of reducing numbers of supermirror elements. 

Pioneer's neutron guide is approximately 50 m long and comprises numerous supermirror elements. Precise alignment of these elements during assembly is crucial to ensuring optimal performance. Manufacturers typically limit individual supermirror lengths to 0.5~m. To minimize misalignment effects, we prioritize the use of 0.5-m elements over shorter ones whenever feasible. This approach contrasts with the ideal elliptical guides, which have continuously varying cross-sections. Although truly curved guides are commercially available, achieving high-precision curved surfaces presents engineering challenges. As a result, Pioneer adopts a polygonal approximation for curved guide sections, favoring the use of shorter supermirror elements. To address potential concerns, we simulated the performance and confirmed that the polygonal approximation, using 0.5-meter supermirror elements, satisfies Pioneer's performance requirements.

Supermirrors are advanced meta-materials consisting of multilayered structures, typically Ni/Ti or MoNi/Ti, deposited on atomically smooth substrates with varying layer thicknesses~\cite{hayter1989discrete}. MoNi/Ti supermirrors are favored to transport polarized beams, where a small percentage of Mo is added into the Ni layer to suppress magnetization, thereby reducing depolarization effects without compromising reflectivity~\cite{schebetov1999multi, padiyath2004influence, kovacs2008non}. Supermirrors are characterized by their $m$-values, which define the critical wavevector transfer, $mQ_{\text{C, Ni}}$, up to which they maintain high neutron reflectivity (typically 50-99\%). $Q_{\text{C, Ni}}$ is the critical moment transfer for natural Ni. Modern manufacturing techniques can produce supermirrors with $m$-values as high as 8~\cite{schanzer2016neutron}; however, this comes with trade-offs, which will be discussed later. Therefore, the $m$-values for supermirrors should be carefully selected based on the specific requirements of neutron transport, avoiding the use of unnecessarily high coatings.

\subsection{Guide Geometry Optimization}{\label{subsec:guide}}
There are three primary objectives for optimizing the guide geometry and beam control system~\cite{liu2024general}: (1) delivering a high-flux beam within the region of interest (ROI) at the sample position to maximize signal, (2) removing unwanted neutrons from the incident beam early to reduce background, and (3) providing a uniform beam at the sample position to minimize systematic errors. To achieve these goals for diverse experimental needs, Pioneer will deploy four slit packages and an exchangeable pre-sample aperture for beam control. A thorough investigation determined the optimal number and placement of these slits. The final configuration includes two in-guide slits: one immediately after the T$_0$ chopper and the other at the shared focal point of two half-elliptical guide sections. Two additional slits are placed downstream of the guide, with a final aperture located just before the sample environment within the reentrant vessel~\cite{liu2025optical}. The aperture's length and opening are customizable according to the sample environment and desired beam size.

Unlike the simple straight-elliptical guide geometry presented in our earlier conceptual study~\cite{liu2024general}, the real guide geometry requires further optimization to accommodate large in-guide gaps and maximize beam control performance. To address these needs, we considered three representative operation scenarios: (1) high flux with a large sample (HF-LG), (2) high flux with a small sample (HF-SM), and (3) high resolution (HR). The spatial and divergence ROIs for these cases are as follows: for HF-LG, a $5 \times 5$~mm$^2$ area with $\pm 0.35^\circ$ divergence in both horizontal and vertical directions; for HF-SM, a $1 \times 1$~mm$^2$ area with $\pm 0.32^\circ$ divergence; and for HR, a $3 \times 3$~mm$^2$ area with $\pm 0.14^\circ$ divergence. For each optimization iteration, three instrument instances were generated with identical guide geometries but different slit and aperture settings.

For Monte Carlo ray tracing simulations and numerical optimization, we utilized the McStas~\cite{willendrup2014mcstas}, MCViNE~\cite{lin2016MCViNE}, SciPy~\cite{2020SciPy-NMeth}, and JupyterLab~\cite{kluyver2016jupyter} software packages. The guide system was modeled using the McStas component {\texttt{Guide\_gravity}} to incorporate gravity effects. A user-defined reflectivity model was applied with an $m$-value of 6, a slope parameter $\alpha = 3.044$~\AA, and a waviness value of 0.0057$^\circ$ ($1 \times 10^{-4}$ rad)~\cite{liu2024general}. The wavelength-dependent specular reflectivity (i.e., $\alpha$), which arises from the local field effect~\cite{kolevatov2019neutron}, was neglected. While an $m$-value of 6 exceeds what are needed, the optimal $m$-values were determined after the guide geometry optimization, as detailed in Sec.~\ref{subsec:supermirror}. Chopper effects were not considered during the guide geometry optimization, and attenuation effects from aluminum windows and air scattering along the beam path were ignored too. 

The transport performance was evaluated within predefined ROIs in the four-dimensional (4D) phase space at the sample location. The optimization metric also included the beam control efficiency of the four slit packages. The metric for an individual instrument instance was defined as, 
\begin{equation}
\begin{aligned}
\text{metric}_\text{i} &= I_\text{roi} - (w_0 \times I_\text{outside-div-roi} + w_1 \times I_\text{outside-spatial-roi}) \\
&\quad - (w_2 \times \delta I_\text{div} + w_3 \times \delta I_\text{spatial}) \\
&\quad - \sum_{s=1}^{4} a_s \times \eta_s,
\end{aligned}
\end{equation}

where all terms except the last one had been considered previously~\cite{liu2024general}. $I_\text{roi}$ is the total flux within ROI, $I_\text{outside-div-roi}$ and $I_\text{outside-spatial-roi}$ represent the undesired flux outside the angular and spatial ROIs, respectively. $\delta I_\text{div}$ and $\delta I_\text{spatial}$ are the standard deviations of the angular and spatial distributions, only applied to the phase space within the spatial ROI. The last term takes into account the slit system, where the slit efficiency $\eta_\text{i}$ is defined as, 
\begin{equation}
\begin{aligned}
\eta_\text{i}~= 1- \frac{I_\text{post-slit} - I_\text{roi}}{I_\text{pre-slit} - I_\text{roi}},
\end{aligned}
\end{equation}
with $I_\text{post-slit}$ and $I_\text{pre-slit}$ being the total flux before and after the slit, respectively. The parameters $a_1$-$a_4$ are scaling factors that incorporate slit efficiency into the metric, whose typical values range from 2\% to 10\% of $I_\text{roi}$. If a slit fails to remove any neutrons,  $I_\text{post-slit} = I_\text{pre-slit}$, resulting in zero efficiency. Conversely, if $I_\text{post-slit}$ matches the flux within the desired ROI, the slit has successfully removed all unwanted neutrons, achieving 100\% efficiency. The aperture efficiency is not included in the metric since its endpoint depends on the specific sample environment. Details regarding the exchangeable aperture system are reported elsewhere~\cite{liu2025optical}. For this analysis, we assumed the aperture ends at 0.20~m from the nominal sample position. The overall metric for the guide geometry optimization combines the contributions of the three instrument instances as, 
\begin{equation}
\begin{aligned}
metric_\text{total}~= \sum_{i=1}^{3} C_\text{i} \times metric_\text{i}.
\end{aligned}
\end{equation}

where $C_\text{i}$ is the weighting factor for individual instance.

The optimization involves 8 hyperparameters ($w_\text{0}$ - $w_\text{4}$, $a_\text{1}$-$a_\text{4}$) for each instrument instance and 3 additional weighting factors (\(C_\text{i}\)) to calculate the combined metric. These hyperparameters are determined empirically using the strategy outlined previously~\cite{liu2024general}. The optimization used a wavelength range of 0.8-8.0~\AA, slightly broader than the instrument's minimal operational wavelength range. Within this range, the influence of gravity is minimal~\cite{liu2024general}. Therefore, a guide with a square cross-section, exhibiting symmetry between horizontal and vertical directions, was considered. 

\begin{figure*}
\includegraphics[width=0.85\textwidth]{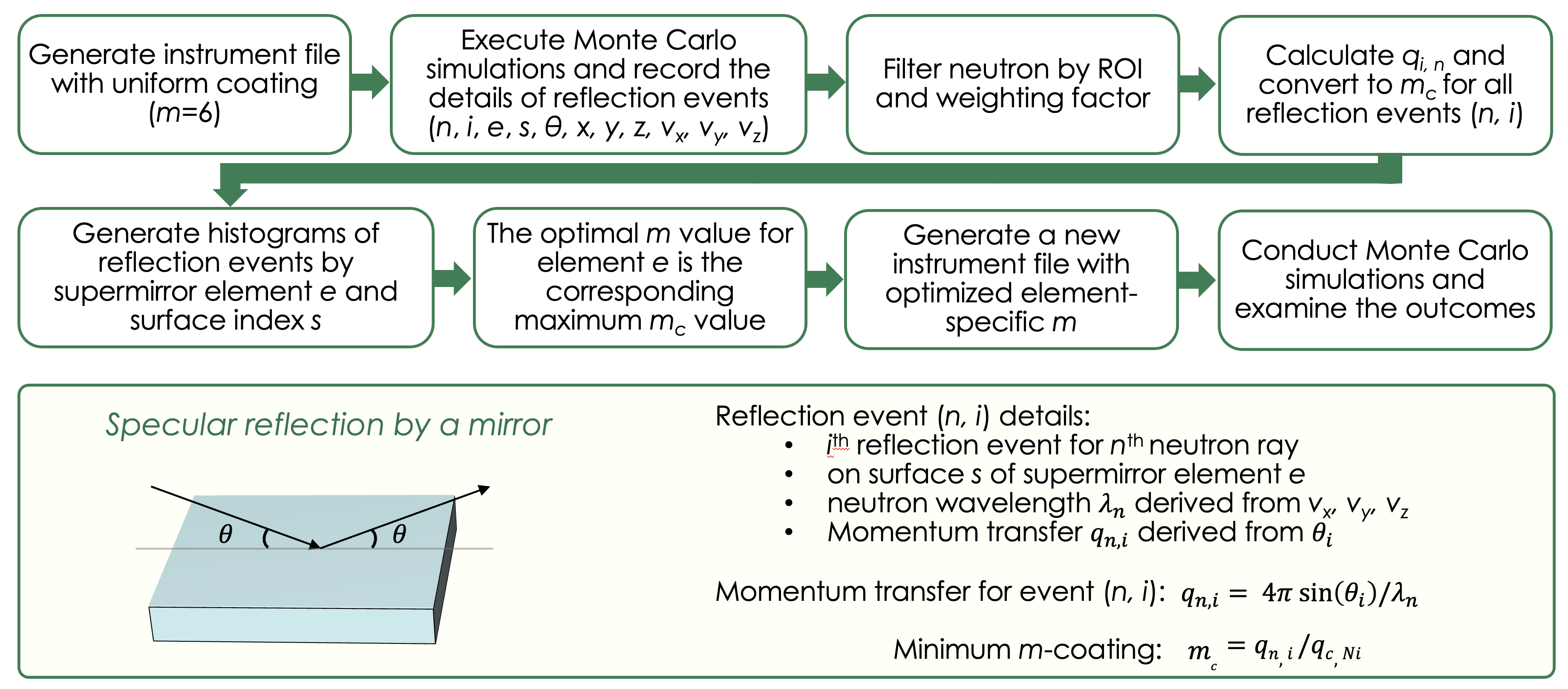}
\caption{\label{fig:mProfileOpt} (top) Workflow to optimize the guide coating. (bottom) Notations and equations of specular reflections that are used to calculated the minimum $m$ value of a supermirror element. }
\end{figure*}

\begin{figure*}[th]
\includegraphics[width=0.85\textwidth]{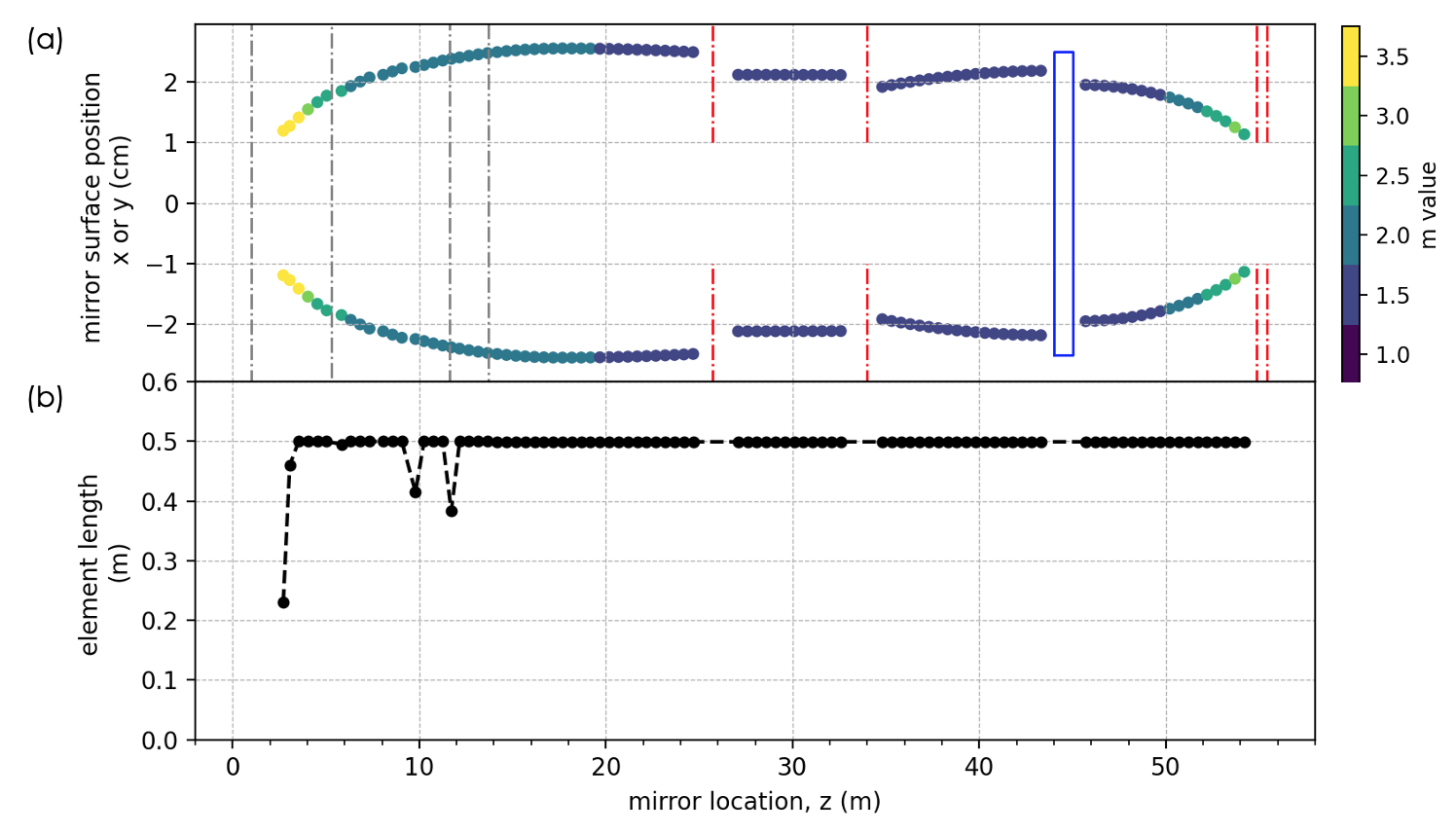}
\caption{\label{fig:supermirrors} (a) Optimized guide geometry and coating profile. Colored dots indicate the positions of the mirror surfaces at the centers of individual elements, with their $m$-values represented by color coding. The guide system is interrupted at the gray lines, which mark the boundaries of the monolith inserts (first pair) and the bunker wall (second pair). Red dashed lines denote the slit positions, and the blue box indicates the location of the polarizing V-cavity. The moderator surface and the nominal sample position are at z = 0.0~m and z = 56.0~m, respectively. (b) Lengths of individual mirror elements.}
\end{figure*}

\subsection{Supermirror Coating Optimization}{\label{subsec:supermirror}}
The optimal coating for individual supermirror element was determined after the guide geometry optimization. The workflow for this process is outlined in Fig.~\ref{fig:mProfileOpt}. The HF-LG instance was selected for this analysis, as it demands the largest phase space, thus defining the most stringent coating requirements. Monte Carlo ray tracing simulations were initially conducted using the optimized guide geometry with a uniform coating profile ($m = 6$). During the simulations, details of reflection events were recorded, including the event index ($i$) of the neutron ray (with an index of $n$), supermirror element index ($e$) and surface index ($s$) (where $s = 1, 2, 3, \text{or}~4$ corresponds to the four side surfaces of the guide), and the reflection angle ($\theta$), the reflection coordinates ($x, y, z$) and the post-reflection velocity components ($v_\text{x}, v_\text{y}, v_\text{z}$). To determine the optimal coating values, only neutrons arriving within the desired phase space ROI at the sample location were considered. Neutron rays with low statistical weight were excluded from the analysis, as they would drive the $m$-values higher without significant performance gains. 

The wavevector transfer, $q_\text{n, i}$, for each reflection event was calculated based on the neutron velocity and reflection angle. The minimum $m$-value, $m_\text{c}$, required for significant reflectivity (\(\mathcal{O}(1)\)) was derived from $q_\text{n, i}$. Given the symmetric guide configuration and negligible gravity effects, the four side surfaces were treated equivalently. All the events were then grouped by their corresponding supermirror element index. The highest $m_\text{c}$ with in each group was determined as the optimal coating value for that element. Finally, the instrument file was updated with the optimized, element-specific $m$-values, for Monte Carlo simulations to evaluate the guide optimization performance.

\subsection{Optimized Guide}{\label{subsec:guideopt}}
\begin{figure}
\includegraphics[width=0.42\textwidth]{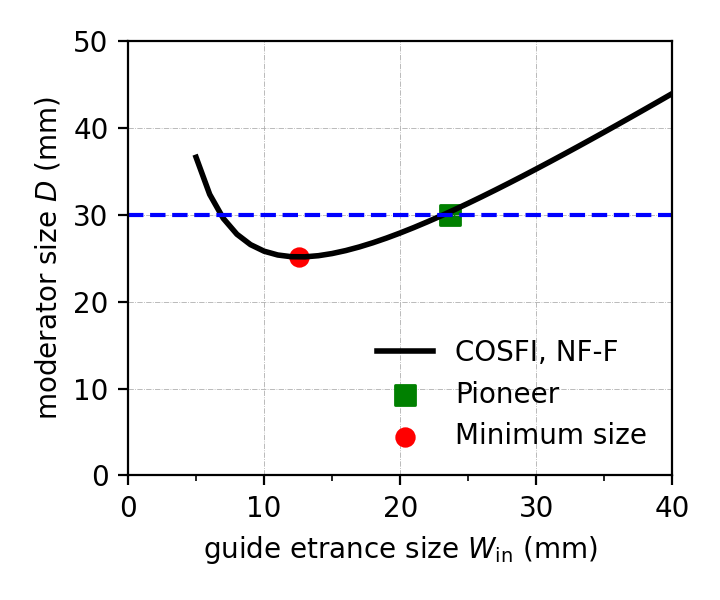}
\caption{\label{fig:COFSI}  The NF-F COFSI function calculated with Pioneer's parameters. Pioneer's optimized guide geometry aligns with the optimal condition that correlates the moderator size and guide entrance size, as indicated by the square green symbol sitting on the COFSI function. The blue dashed line shows the STS moderator size of 30~mm, while the red circle marks the case corresponding to the minimum moderator size.}
\end{figure}

\begin{table}
\caption{\label{tab:slits} Requirements of three selected scenarios used for the guide geometry optimization. Also shown are the optimization results, including the slits settings, the flux evaluated over spatial ROIs within the wavelength band of 0.8-5.1~\AA, and the maximum flux peaked at 2.5~\AA.}
\centering
\begin{tabular}{>{\raggedright\arraybackslash}p{3.0cm}|>{\centering\arraybackslash}p{1.68cm}|>{\centering\arraybackslash}p{1.68cm}|>{\centering\arraybackslash}p{1.68cm}}
\hline \hline
\textbf{Name}                        & \textbf{HF-LG} & \textbf{HF-SM}     & \textbf{HR}                      \\ \hline
beam size (mm)                       & 5.0            & 1.0                & 3.0 \\ \hline
beam divergence                      &0.70$^\circ$    &0.65$^\circ$        &0.28$^\circ$ \\ \hline
integrated flux                      & & & \\ 
\hspace{2pt}[$10^8~\text{n}/(\text{cm}^{2}\cdot \text{s})$]              &12.3      &9.9                & 1.8 \\ \hline
peak flux                            & & & \\ 
\hspace{2pt}[$10^8~\text{n}/(\text{\AA}~\cdot \text{cm}^{2}\cdot \text{s})$]    &5.8      &4.7                & 0.87 \\ \hline
slit 1 (cm)                         &4.65             &3.75     &4.15 \\ \hline
slit 2 (cm)                         &4.08             &4.00     &4.03 \\ \hline
slit 3 (cm)                         &1.98             &1.63     &0.69 \\ \hline
slit 4 (cm)                         & 1.20            &0.80     &0.72 \\ \hline
aperture (cm)                       & 0.69            &0.28     &0.33 \\ \hline
\end{tabular}
\end{table}

Figure~\ref{fig:supermirrors}(a) presents the optimized guide geometry and coating profile. The guide system consists of two half-elliptical guides with a shared focal point, connected by three nonlinearly tapered sections. The first half-elliptical guide starts at 2.59 m and ends at 17.43 m, and the second one starts from 45.46 m and ends at 53.96 m. Supermirror elements are 0.5~m in length, except for those near breaking points caused by the monolith and bunker walls. 


Notably, the first half-elliptical guide section contains six subsections~\cite{liu2025optical}. The first three gaps are introduced by the monolith wall and two disk choppers, as shown in Fig.\ref{fig:supermirrors}(a). The last two gaps, resulting from the bunker wall, are narrower but affect the lengths of the supermirror elements, as seen from Fig.\ref{fig:supermirrors}(b). These six elliptical subsections share common focal points, maintaining the same long axis while their short axes progressively increase towards the moderator. A point source approximation was applied to establish the relative values of the six short axes, ensuring that the first half-elliptical guide views the moderator without introducing divergence gaps. This approach reduces the number of free parameters. Independent optimization of the short-axis values for each subsection did not yield significant performance improvements, validating the effectiveness of our simplified approach.

The optimization result can be assessed using Curves of Optimal and Full Sample Illumination (COFSI) functions formulated by Konik and Ioffe~\cite{konik2023new}. COFSI functions analytically describe the optimal relationship between moderator size, neutron transport geometry, and phase space requirements at the sample position. They provide a concise yet valuable measure for assessing instrument design effectiveness in maximizing sample flux, minimizing background, and reducing shielding costs along the beamline. The form of the COFSI function depends on the characteristics of the guide at the entrance and exit: focusing (F) or non-focusing (NF). Konik and Ioffe conducted Monte Carlo simulations for realistic guide systems and found that elliptical guides closely resemble the NF-F configuration, where the entrance exhibits non-focusing characteristics and the exit demonstrates focusing properties.

For the NF-F configuration, the COFSI function is given by $D_{\text{mod}} = \frac{2 d_\text{s} \alpha_\text{s} L_{\text{in}}}{W_{\text{in}}} + W_{\text{in}}$, where $D_{\text{mod}}$ is the moderator size, $L_{\text{in}}$ is the distance between the moderator and the guide entrance, $W_{\text{in}}$ is the width of the guide entrance, and $d_\text{s}$ and $\alpha_\text{s}$ are the required beam size and a half of the beam divergence at the sample position. 

Figure~\ref{fig:COFSI} shows the NF-F COFSI function calculated with Pioneer's parameters, where $d_\text{s} = 5$~mm and $\alpha_s = 0.35^\circ$, representing the largest requested phase space volume. The guide optimization results in $L_{\text{in}} = 2.59$~m and $W_{\text{in}} = 23.75$~mm, while the view size of the STS moderator size is 30~mm. As indicated by the green square marker, Pioneer's optimized guide aligns with the optimal condition specified by the COFSI function. 

The optimized coating profile, color-coded in Fig.~\ref{fig:supermirrors}(a), exhibits  $m$-values ranging from 1.5 to 3.5, rounded up to the nearest half-integer to simplify engineering. As expected,  higher $m$-values are observed at the narrower ends of the elliptical guide sections, where the maximum reflection angles are greater. 

\begin{figure*}
\includegraphics[width=1.0\textwidth]{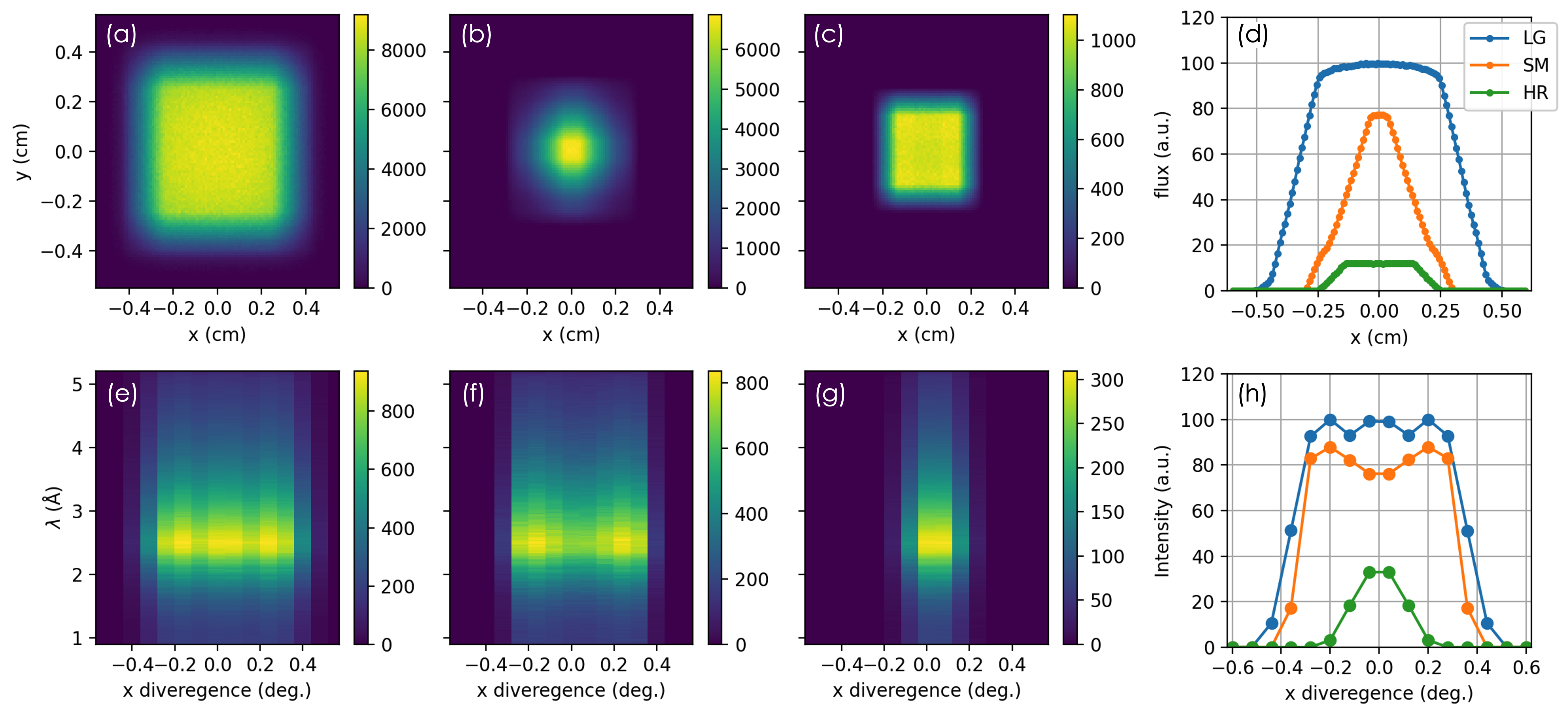}
\caption{\label{fig:beamprofiles} Beam profiles at the sample positions for the three selected cases, HF-LG ($5 \times 5$~mm$^{2}$, $0.70^\circ \times 0.70^\circ$), HF-SM ($1 \times 1$~mm$^{2}$, $0.65^\circ \times 0.65^\circ$) and HR ($3 \times 3$~mm$^{2}$ $0.28^\circ \times 0.28^\circ$). (a-c) Spatial beam profiles for HF-LG, HF-SM, and HR, respectively. (d) Line profiles of the spatially resolved fluxes along the horizontal ($x$) direction, with the signals integrated along the vertical direction within their respective ROIs. (e-g) Wavelength-resolved angular distributions of the beam profiles for HF-LG, HF-SM, and HR, respectively, with signals integrated over the corresponding spatial ROIs. (h) Line profiles of the angular-resolved beam intensities along the horizontal direction, with signals integrated over the respective spatial ROIs.}
\end{figure*}

\begin{figure}
\includegraphics[width=0.45\textwidth]{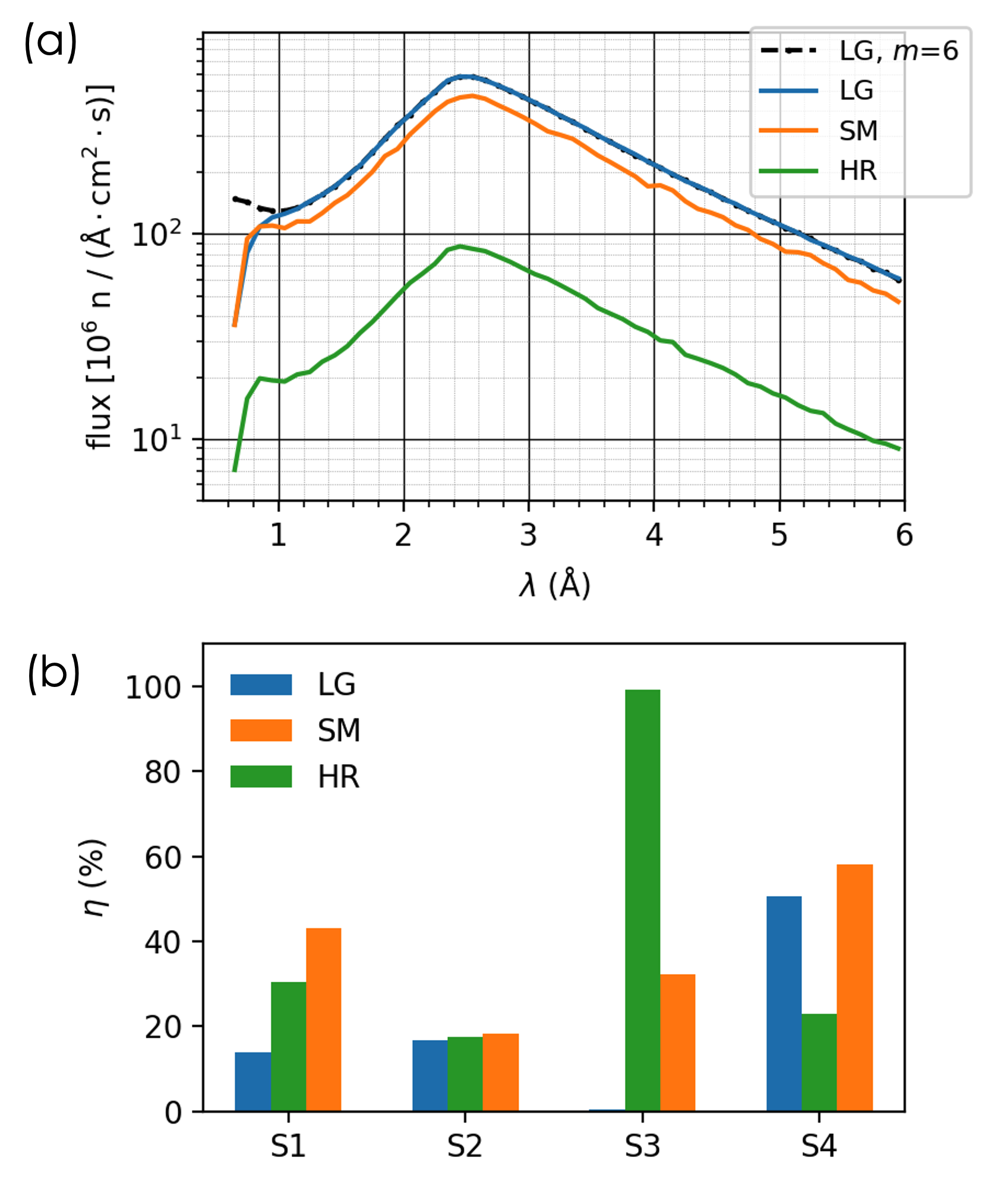}
\caption{\label{fig:beamcontrol}  (a) Fluxes for the three selected cases, evaluated over their respective spatial ROIs. Also shown is the flux from the HF-LG case before applying the optimized guide coating. (b) Slit system effectiveness in removing unwanted neutrons from the incident beam.}
\end{figure}

\begin{figure}[htb]
\includegraphics[width=0.45\textwidth]{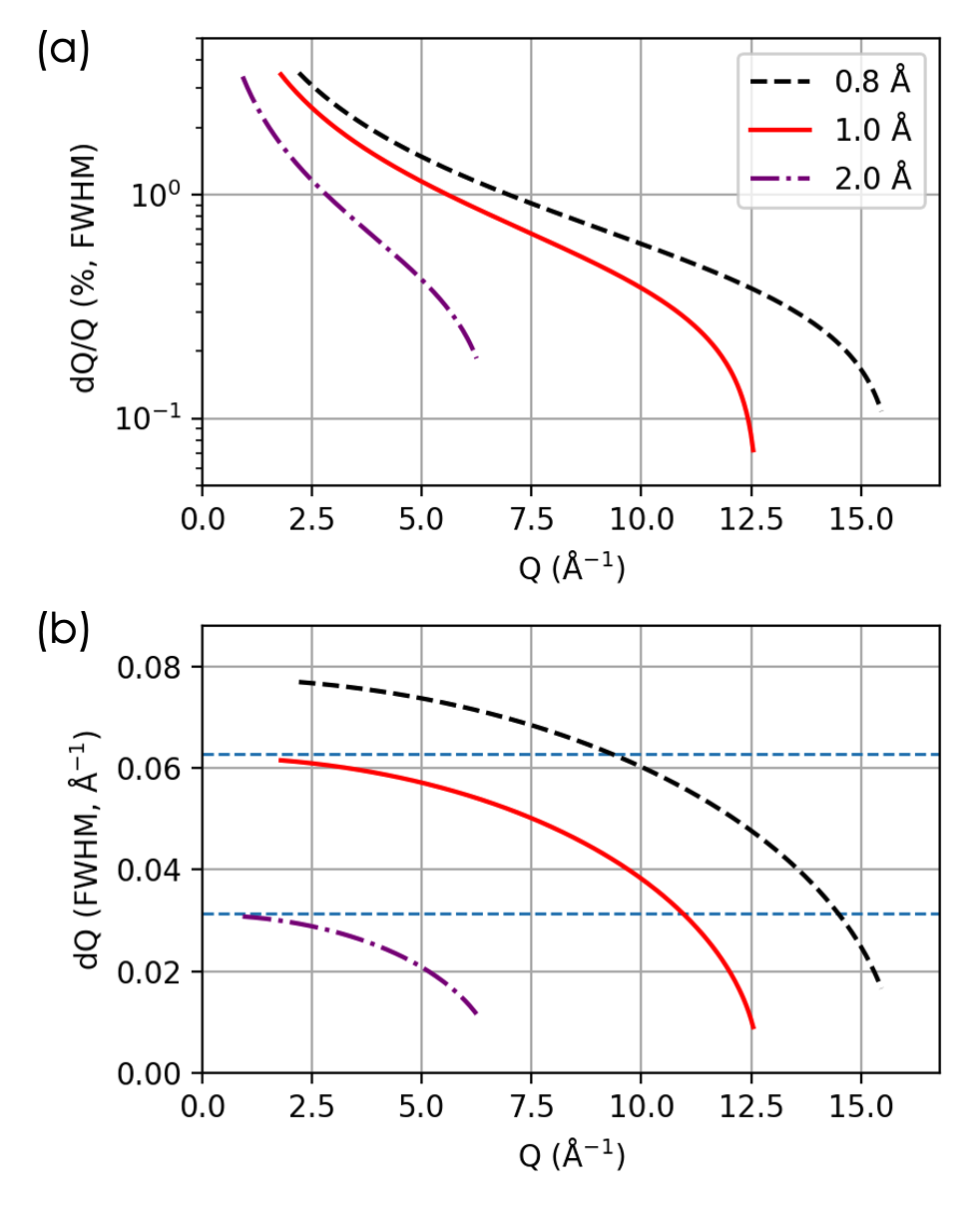}
\caption{\label{fig:HR}  (a) Relative instrument resolution $\text{d}Q_\text{instr}/Q$ and (b) Absolute instrument resolution $\text{d}Q_\text{instr}$ calculated with $\text{d}\theta_\text{beam}=0.28^{\circ}$ (the HR case) as a function of the momentum transfer $Q$ for three selected wavelengths. The dashed lines in (b) denote the constant $\text{d}Q$ values of $6.28 \times 10^{-2}$~\AA$^{-1}$ and $3.14 \times 10^{-2}$~\AA$^{-1}$. }
\end{figure}

\subsection{Beam Transport and Control Performance}{\label{subsec:transport}}
Figure~\ref{fig:beamprofiles} presents the beam profiles at the sample position generated from the optimized guide geometry, using three distinct slit settings corresponding to HF-LG, HF-SM, and HR configurations. Details of the slit settings are provided in Tab.~\ref{tab:slits}. The maximum wavelength-integrated fluxes on samples in the high-flux modes are $\sim 1 \times 10^9~\text{n}/(\text{cm}^{2}\cdot \text{s})$, evaluated from a full band of 0.8-5.1~\AA. 

The spatial-resolved profiles, shown in Figs.~\ref{fig:beamprofiles} (a)-(c), exhibit high homogeneity within ROIs. Figure~\ref{fig:beamprofiles} (d) displays horizontal line cuts of the spatially resolved flux, integrated along the vertical direction within their respective ROIs. The flat-top widths of these profiles correspond to the required homogeneous beam regions for each instance, as listed in Tab.~\ref{tab:slits}. Figures~\ref{fig:beamprofiles} (e)-(g) show the wavelength-resolved angular distributions, with signals integrated over the corresponding spatial ROIs. Notably, the beam divergences remain nearly wavelength-independent, a feature ensuring spatially homogeneous flux within the ROIs. Panel (h) displays horizontal line cuts of the angular-resolved flux, integrated over the wavelength range of interest, from which the beam divergences are determined.  The results demonstrate that the neutron transport and control systems deliver uniform and configurable beam profiles, accommodating spatial ROIs from 5~mm to 1~mm and beam divergences from 0.70$^\circ$ to 0.28$^\circ$. 

Figure~\ref{fig:beamcontrol}(a) displays the wavelength-resolved flux for the three instances. The spectra peak at 2.5~\AA, consistent with the source brilliance, due to the wavelength-insensitive neutron transport efficiency~\cite{liu2024general}. This aligns with the observed wavelength-independent beam divergence. Given the homogeneous beam profiles, the three spectra exhibit similar wavelength dependencies, differing primarily by constant factors. The guide coating is optimized for a cutoff wavelength $\lambda_{\text{cutoff}}$ of 1.0~\AA. Below this threshold, the flux profiles before and after optimization diverge. Although Pioneer can transport shorter-wavelength neutrons, the transport efficiency drops rapidly below 0.8~\AA, effectively setting the operational cutoff wavelength.

The slit systems also control the beam divergence, enabling high-resolution modes. Based on the elastic scattering approximation, the instrument momentum-transfer resolution $\text{d}Q_\text{instr}$ can be estimated as~\cite{schultz2005conceptual},

\begin{equation}
(\frac{\text{d}Q_\text{instr}}{Q})^2 = ( \frac{\text{d}\lambda}{\lambda})^2 + (\frac{\text{d}L}{L})^2  + (\text{cot}(\theta)\text{d}\theta)^2, 
\end{equation}
where the momentum transfer $Q = 4 \pi \text{sin}(\theta)/\lambda$, $\text{d}\lambda / \lambda$ is the relative wavelength uncertainty, $\text{d}L/L$ is the relative flight path uncertainty (which is negligible), and $\text{d}\theta$ is the angle uncertainty. Neglecting sample effects, $\text{d}\theta$ is determined by the beam divergence and detector angular resolution:
\begin{equation}
(\text{d}\theta)^2 = (\text{d}\theta_\text{beam})^2 + (d\theta_\text{det})^2. 
\end{equation}

At low scattering angles, $\text{d}Q_\text{instr}$ is primarily influenced by the angular resolution; while at high scattering angles, $\text{d}Q_\text{instr}$ is limited by the wavelength resolution. Figure~\ref{fig:HR} shows $\text{d}Q_\text{instr}$ estimated with $d\theta_\text{beam}$ = 4.9 mrad (0.28$^\circ$) from the HR mode  and $d\theta_\text{det}$ = 0.75~mrad, derived from the detector spatial resolution of 0.6~mm and the sample-to-detector distance of 800~mm. With $\lambda_{min}$ = 1.0~\AA, $\text{d}Q_\text{instr} < 6.28 \times 10^{-2}$~\AA$^{-1}$ and $Q_\text{max} = 12.5$~\AA$^{-1}$, corresponding to a real-space length scale of 100~\AA~with a resolution of 0.5~\AA. Figure~\ref{fig:VE_QRes} shows the the diffraction patterns from an virtual experiment conducted with a simple-cubic crystal with a lattice constant of 100~\AA~in the HR mode. To probe larger unit cells, $\lambda_{min}$ can be increased to 2.0~\AA,  where $\text{d}Q_\text{instr} < 3.14 \times 10^{-2}$~\AA$^{-1}$ and $Q_\text{max} = 6.25$~\AA$^{-1}$, corresponding to a real-space length scale of 200~\AA~and a resolution of 1.0~\AA. When higher real-space resolutions are needed, $\lambda_{min}$ can be reduced to 0.8~\AA, yielding $Q_\text{max} = 15.6$~\AA$^{-1}$.

\section{Incident Beam Polarizer}{\label{sec:pol}}
Pioneer requires a polarized incident beam option. The performance of a polarizer can be assessed by the parameter $\text{FOM}_\text{pol}$, defined as~\cite{batz20053he}:
\begin{equation}
\text{FOM}_\text{pol} = P^{2}T, 
\end{equation}
where $P$ is the polarization of the neutron beam after the polarizer and $T$ represents the transmission of the polarizer. Polarizing mirrors and $^3$He filters are the most common choices to implement wide-band neutron polarizers. In this section, we will compare these two options per Pioneer's requirements-a polarized incident beam with $\lambda_{min} = 1.2$~\AA, with high FOMs over a broad wavelength band of 4.3~\AA. 

\subsection{Multichannel Polarizing V-cavity}
\begin{figure}
\includegraphics[width=0.49\textwidth]{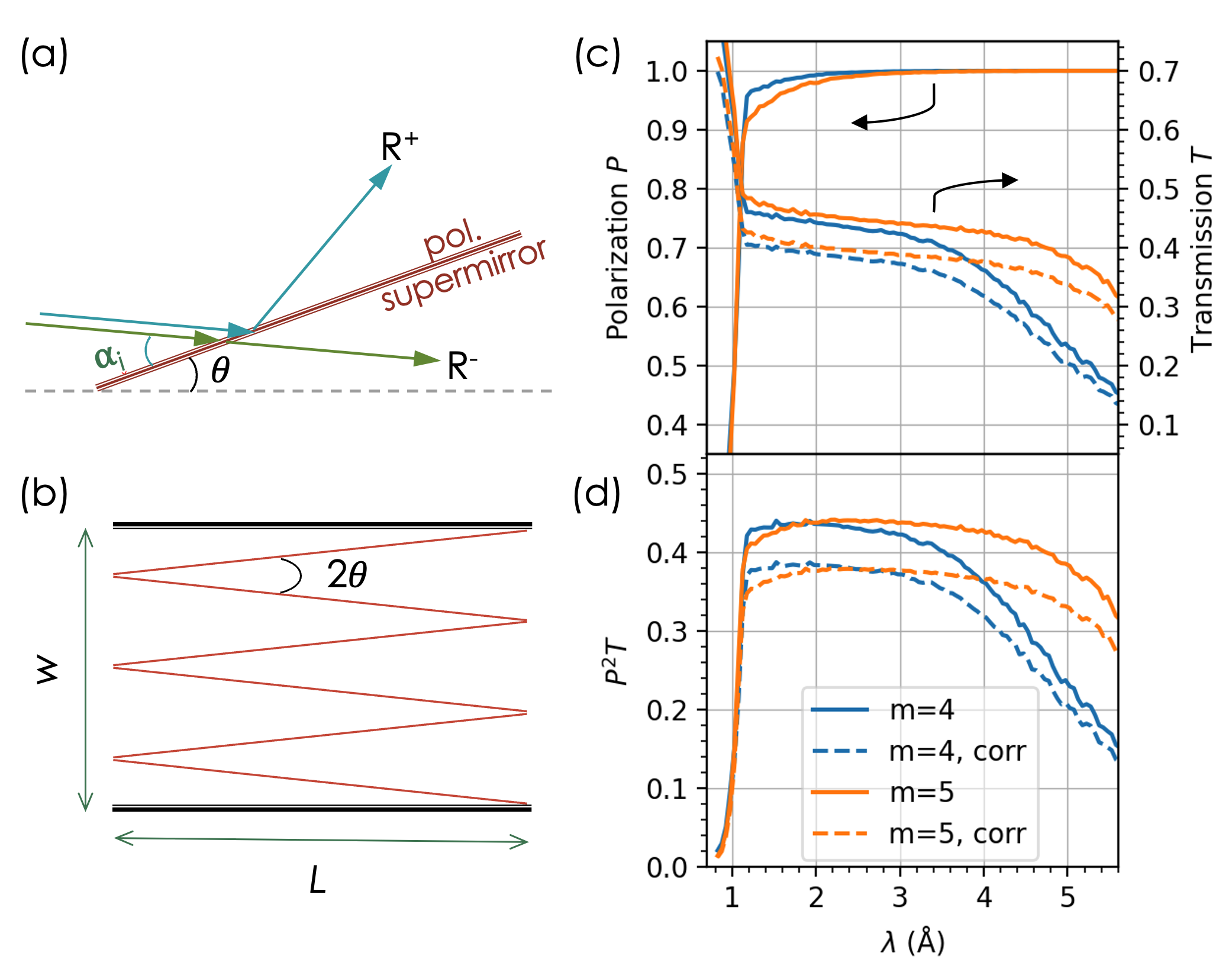}
\caption{\label{fig:V-cavity} (a) Illustration of the principle of a polarizing supermirror. The dashed line denotes the nominal incident beam direction. $\alpha_\text{i}$ and $\theta$ are the glancing incident angle and the tilting angle of the supermirror surface, respectively. (b) Schematic of a multichannel V-cavity with three V channels and six polarizing supermirrors, forming a V-angle of 2$\theta$. The thin dark-red lines indicate the locations of the double-sided polarizing supermirrors, and the thick black lines represent the housings. (c) Polarization and transmission for two V-cavities designed with $m = 4$ and $m = 5$ polarizing supermirrors, respectively, for a cutoff wavelength $\lambda_{cutoff}$ = 1.2~\AA. (d) $P^2T$ for the polarizer as a function of a wavelength. The transmission $T$ and FOM $P^2T$ after accounting for additional transmission loss are also presented.}
\end{figure}

\begin{figure*}
\includegraphics[width=0.8\textwidth]{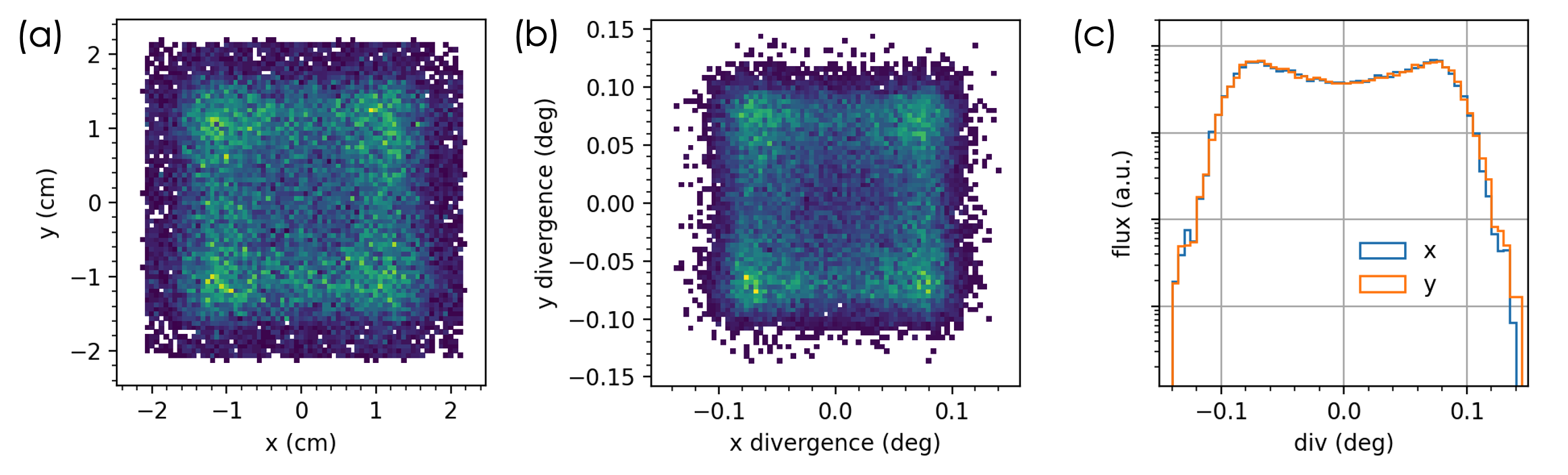}
\caption{\label{fig:beamprofile_pol} (a) Spatial distribution and (b) angular distribution of neutron rays incident on the polarizer. (C) Angular distributions of the neutron beam along the horizontal ($x$) and vertical ($y$) directions. The simulations were based on the HF-LG case, and only neutron rays arriving within the $5\times5$~mm$^2$ spatial ROI at the sample position were included.}
\end{figure*}

Polarizing supermirrors exhibit spin-dependent reflectivity at glancing incidence angles, making them suitable for generating polarized neutron beams~\cite{mezei1976novel}. Fe/Si polarizing supermirrors with high reflectivity up to $m = 5.5$ are commercially available~\cite{schanzer2016neutron}. 

Figure~\ref{fig:V-cavity}(a) illustrates the working principle of a polarizing supermirror. When operating in transmission mode, the transmitted polarized beam maintains the same direction as the incident beam. To ensure a polarizing supermirror selectively transmits a neutron beam of one spin state (down, $-$) and reflects the other (up, $+$) within the wavelength band of [$\lambda_{\text{min}}$, $\lambda_{\text{max}}$] and glancing incident angle [$\alpha_\text{i, min}$, $\alpha_\text{i, max}$], the following conditions shall be satisfied~\cite{stahn2017efficient}, 
\begin{equation}
Q_\text{min}^{-} = \frac{4\pi \sin \alpha_{\text{i, min}}}{\lambda_{\text{max}}} \geq Q_\text{C}^{-}, 
\end{equation}
and 
\begin{equation}
Q_\text{max}^{+} = \frac{4\pi \sin \alpha_{\text{i, max}}}{\lambda_{\text{min}}}  \le Q_\text{C}^{+},  
\end{equation}
where $Q_\text{C}^{-} = Q_\text{Si}$ and $Q_\text{C}^{+} = mQ_\text{C}^\text{Ni}$ are the critical momentum transfers for the down and up spin states,  respectively, assuming a Si substrate. Let $\Delta\alpha = \alpha_\text{i, max} - \alpha_\text{i, min}$ and $\Delta\lambda = \lambda_\text{max} - \lambda_\text{min}$. These conditions leads to, 
\begin{equation}
\alpha_{i, \text{min}} \geq \arcsin \frac{Q_{Si}(\lambda_\text{min} + \Delta\lambda)}{4\pi},
\end{equation}
and 
\begin{equation}
m \geq \frac{4\pi \sin(\alpha_{\text{i, min}} + \Delta\alpha) }{Q_{C}^\text{Ni}\lambda_{\text{min}}}.
\end{equation} 

These requirements impose constraints on the use of polarizing supermirrors for cases with large beam divergence, broad wavelength ranges, or space limitations that prevent achieving small incident angles. Consequently, careful consideration is crucial when positioning the polarizer. Firstly, it should be positioned where the beam divergence is low to minimize the required $m$-values, thereby enhancing performance and reducing engineering complexity. Secondly, it is advantageous to position it where active neutron guides are not critical for the non-polarized beam transport, thereby avoiding the need for translatable guides. An in-guide gap is reserved at Pioneer to accommodate a polarizer up to 1.20~m long. Finally, as the polarization mechanism depends on reflecting neutron rays with undesired spin states out of the beam path, the polarizer shall be located sufficiently far from the detector minimize background from scattered neutrons and enable effective shielding. Considering these factors, we chose to position the polarizer immediately before the second half-elliptical guide~\cite{liu2025optical}, as illustrated in Fig.\ref{fig:supermirrors}(a). 

The polarizer design depends critically on the beam profile at its location. Figure~\ref{fig:beamprofile_pol} shows the spatial and angular distributions of the beam profile along the horizontal ($x$) and vertical ($y$) directions. Only neutron rays within the largest spatial ROI at the sample location and with significant weight are considered. Including all neutron rays striking the polarizer would impose unnecessarily restrictive requirements. The beam spans spatially from -2.13~cm to 2.13~cm in both $x$ and $y$ directions. As seen in Fig.~\ref{fig:beamprofile_pol}(c), the beam exhibits slightly higher divergence along the $y$-direction compared to the $x$-direction due to gravitational effects. Consequently, vertical polarizing mirrors were selected, benefiting from a slightly reduced coating requirements. This choice aligns the beam polarization direction vertically, consistent with the planned magnetic sample environments.

At Pioneer, $\Delta\alpha_\text{x} = 0.28^{\circ}$. Based on the targeted $\lambda_{min} = 1.2$~\AA~and $\lambda_{max} = 5.5$~\AA, the minimum incident angle $\alpha_\text{i,c} = 0.26^{\circ}$, and the minimum coating value $m_\text{c} = 4.5$. This results in a minimum tilting angle $\theta_\text{c} = \alpha_\text{i,c} + \Delta\alpha_\text{x}/2 = 0.40^{\circ}$, relative to the vertical plane. We chose an oversized supermirror width of 4.4~cm to cover the beam cross-section and accommodate misalignment tolerance. This corresponding to a supermirror length of 6.3~m, exceeding the reserved gap of 1.2~m. To address this, a three-channel transmission V-cavity, as illustrated in Fig.\ref{fig:V-cavity}(b), will be used at Pioneer.

The performance of the polarizing V-cavity was evaluated using Monte Carlo ray tracing simulations,  which have used  two multi-V channel components modified from \texttt{Pol\_guide\_vmirror} (by Peter Willendrup) and \texttt{Transmission\_V\_polarisator} (by Andreas Ostermann), respectively. The modified components incorporate multi-V channels, consider transmission loss based on the flight path in polarizing mirrors, and use tabulated reflectivity. Two new components yield essentially identical results, as expected since they describe the same physics, albeit with different implementations. This agreement validates our approach. The used spin-dependent reflectivity of double-side coated Fe/Si polarizing supermirrors on 300-$\mu$m thick Si substrates is shown in Fig.~\ref{fig:PolRef}.  Beam attenuation due to incoherent scattering and absorption by the Si wafer and Fe/Si multilayers was included, with the Si wafer contributing the dominant effect. The beam path length correction caused by the refraction effects is negligible (\(\mathcal{O}(1e-3)\)) and was therefore was ignored. Wavelength-dependent reflectivity effects arising from local field corrections\cite{kolevatov2019neutron, dijulio2022measurements} were neglected, as was the case for the guide geometry optimization. Following simulations, scale factors were applied to account for additional transmission loss arising from the multilayers. 

Two cavities were simulated: one using $m = 5$ polarizing supermirrors and the other using $m = 4$, as reflectivity data for only integer-$m$ polarizing supermirrors were available. The cavity housing's inner surfaces were assumed to have a perfectly absorptive coating layer ($m = 0$). For $m = 5$, the V-cavity has a V-angle $2\theta = 0.90^{\circ}$ and a length of 0.93~m. In contrast, $m = 4$ necessitates the smallest V-angle $2\theta = 0.70^{\circ}$ permitted by the reserved gap size, resulting in a length of 1.20~m. For the $m = 4$ case, the minimum incident angle $\alpha_\text{i,min} = 0.21^{\circ}$, falling below $\alpha_\text{i,c} = 0.26^{\circ}$, leading to significantly reduced transmission efficiency for longer-wavelength neutrons, as confirmed by simulations. Figures~\ref{fig:V-cavity}(c) and (d) show the polarization $P$, transmission $T$ and FOM $P^2T$ for the $m = 4$ and $m = 5$ cavities. While both achieve a cutoff wavelength $\lambda_\text{cutoff}$ of 1.2~\AA~with high polarization ($P > 90\%$), $P^2T$ for $m = 4$ decreases significantly for $\lambda > 4.0$~\AA, whereas $m = 5$ extends the high-FOM band by 1.3~\AA. 

Previous experiments with $m = 3$ and $m = 4$ Ni/TiMo supermirrors reported 6\% beam attenuation for beams transmitted through the supermirror multilayers above the critical momentum transfer\cite{dijulio2022measurements}, attributed to various scattering processes. Higher-$m$ supermirrors are expected to exhibit increased attenuation. Assuming Fe/Si multilayers exhibit similar attenuation levels, scale factors of 88\% and 86\% were applied for $m = 4$ and $m = 5$ double-side coated Fe/Si supermirrors, respectively. The corrected transmissions and FOMs are presented in Figs.\ref{fig:V-cavity}(c) and (d), showing an average FOMs of 32\% and 36\% over the wavelength range of 1.2-5.5~\AA~for the $m = 4$ and $m = 5$ cavities, respectively.

\subsection{$^3$He spin filter}
\begin{figure*}
\includegraphics[width=0.96\textwidth]{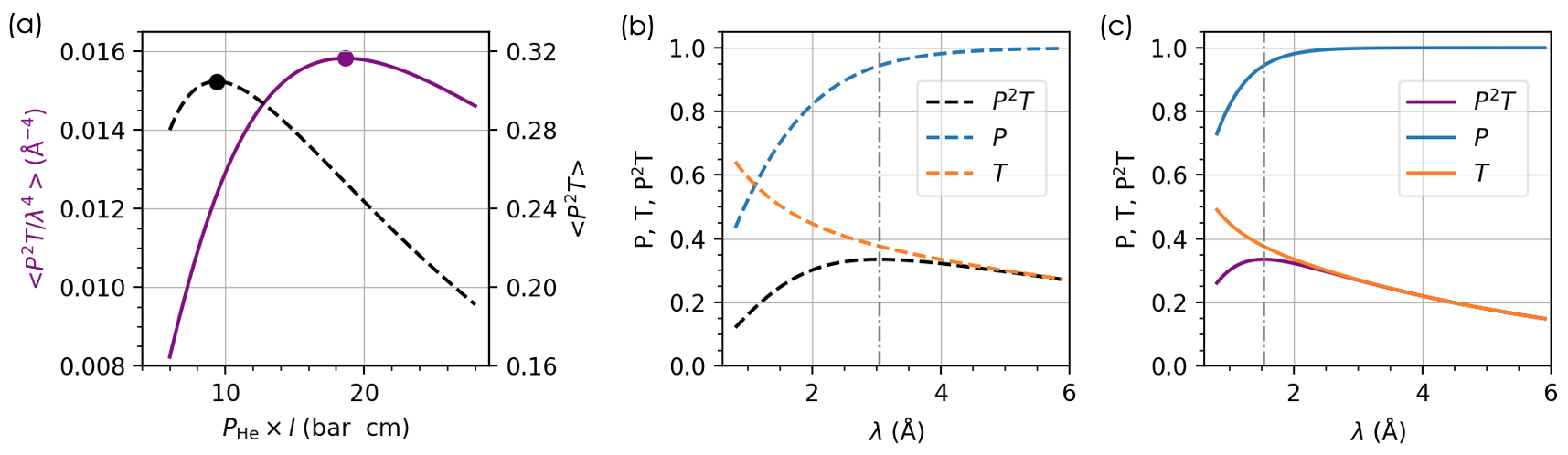}
\caption{\label{fig:3He} Performance of $^3$He spin filters, assuming $P_\text{He}$ = 85\%. (a)  The dashed black line and the solid purple line represent the FOM $P^{2}T$ and the effective FOM $P^{2}T/\lambda^{4}$, respectively, as functions of the product of the gas pressure and the length of the $^3$He cell, $p_\text{He} \cdot l$, averaged over the wavelength range of $1.2-5.5$~\AA. The $p_\text{He} \cdot l$ values corresponding to the maximum average FOM and maximum average effective FOM are 9.36 and 18.64~$\text{bar} \cdot \text{cm}$, respectively. (b) and (c) Wavelength-dependent polarization parameters $P$, $T$, and $P^{2}T$ of $^3$He spin filters for $p_\text{He} \cdot l$ values corresponding to the maximum average FOM and maximum average effective FOM, respectively. The dot-dashed lines indicates the wavelengths with the peak $P^{2}T$ values. }
\end{figure*}

A $^3$He neutron spin filter polarizes neutron beams by exploiting the spin-dependent absorption cross-section of nuclear spin-polarized $^3$He. Neutron absorption predominantly occurs when the spins of the neutron and $^3$He nucleus are anti-aligned. The absorption cross-section with spin antiparallel to the $^3$He nuclear spin is 10666 barns for 25~meV neutrons, while for the parallel case the scattering cross-section is only a few barns~\cite{batz20053he, gentile2017optically}. An ideal $^3$He spin filter would absorb nearly all neutrons with antiparallel spins while transmitting most neutrons with parallel spins. This approach eliminates issues with finite beam divergences and background from undesired spin states. $^3$He spin filters are compact and work over a broad energy range compared to polarizing mirrors. However, they require either an in-situ or an ex-situ laser optical pumping system. The highest reported $^3$He polarization values for neutron spin filters range from 75\%-85\%~\cite{ye2013wide, chen2014limits, salhi2019situ}. 

Two critical parameters characterize the effectiveness of a $^3$He spin filter: transmission $T$ and opacity $O$. Transmission $T$ for neutrons with spins parallel ($T_{+}$) and antiparallel ($T_{-}$) to the $^3$He spin can be calculated as:
\begin{equation}
\begin{split}
T_{\pm} & =  \exp{(-(1 \mp P_\text{He})n_\text{He}l\sigma_0)} \\
        & =  \exp{[-O(1\mp P_\text{He})]},
\end{split}
\end{equation}
where $P_{He}$ is the $^3$He nuclear polarization, $n_{He}$ is the number density of $^3$He atoms, $l$ is the cell length and $\sigma_0$ is absorption cross-section for unpolarized neutrons ( $\approx 3000 \cdot \lambda$ [\AA] barns). Opacity ($O = n_\text{He}l\sigma_0)$) quantifies the filter's absorption efficiency. The conventional opacity $O'$ is defined based on the gas pressure $p_\text{He}$~\cite{batz20053he} as, 
\begin{equation}
O' = p_\text{He} [\text{bar}] \cdot \lambda [\text{\AA}] \cdot l [\text{cm}], 
\quad O = 0.0732 O'.
\end{equation} 
Neutron polarization after passing through the spin filter is given by:
\begin{equation}
P_n = \frac{T_+ - T_-}{T_+ + T_-} = \tanh(OP_\text{He}).
\end{equation} 

For a given $P_\text{He}$, the performance of a $^3$He spin filter is primarily determined by its opacity, which exhibits strong wavelength dependence. Consequently, the product of gas pressure and the cell length ($p_\text{He} \cdot l$) must be optimized per the desired wavelength band. Figure~\ref{fig:3He}(a) presents the calculated FOM $P^2T$ of $^3$He spin filters as a function of $p_\text{He} \cdot l$, averaged over $1.2-5.5$~\AA. The calculation assumes an optimistic $^{3}$He polarization value $P_{He}$ = 85\%. However, considering the impact of magnetic form factors and the wavelength-dependent reciprocal space coverage ($\text{d}V_\text{Q} = \text{d}\lambda / \lambda^4$), high-Q peaks collected with shorter wavelengths will significantly affect the required counting time. To account for this, we introduced an effective FOM ($FOM_\text{pol, eff} = P^2T/\lambda^n$) to prioritize shorter wavelength neutrons. For a series of peaks collected at the same angle at a time-of-flight single-crystal diffractometer,  the integrated peak intensity is proportional to $\lambda^4$, after ignoring the differences in the structure factors, and the wavelength-dependence of flux, detector efficiency, sample absorption and secondary extinction correction~\cite{buras1975relations}. Therefore, we selected $n=4$ for this analysis. The optimal $p_\text{He} \cdot l$ values corresponding to the maximum FOM and maximum $\text{FOM}_\text{eff}$ are 9.36 and 18.64 $\text{bar} \cdot \text{cm}$, respectively.  

Figure~\ref{fig:3He}(b) displays the wavelength-resolved polarization parameters $P$, $T$, and $P^{2}T$ for the optimal condition maximizing the average FOM over $1.2-5.5$~\AA. The average FOM is 0.30, with $P^{2}T_\text{max} = 0.34$ at 3.04~\AA. However, $P^{2}T$ is only 0.20 at 1.2~\AA. Figure~\ref{fig:3He}(c) shows the results for the condition maximizing $\text{FOM}_\text{eff}$, where the average FOM is 0.16, with $P^{2}T_\text{max} = 0.34$ at 1.52~\AA, and $P^{2}T$ increases to 0.32 at 1.2~\AA. $^3$He cell walls typically introduce an additional $\sim$ 12\% beam loss~\cite{gentile2017optically}. Applying a scale factor of 0.88 to the transmission, the average FOMs becomes 0.27 and 0.14 for the two respective cases.

\section{Discussion}{\label{sec:discussion}}
Earlier studies demonstrated that using a smaller para-hydrogen moderator can enhance neutron source brilliance~\cite{zhao2013optimizing, batkov2013unperturbed}. Figure~\ref{fig:COFSI} indicates that the minimum optimal moderator size is $D_{\text{opt}} = 25.2$~mm, smaller than the STS moderator size, suggesting a potential performance gain for Pioneer with a reduced moderator size.  However, Monte Carlo simulations offer a more realistic assessment than the COFSI function, as they account for factors such as non-ideal reflectivity and specific guide geometries. Simulations conducted by Konik and Ioffes highlight two key deviations from analytical predictions: (1) wavelength-dependent effects due to imperfect supermirror reflections and (2) a shift in the optimal configuration toward a slightly larger moderator and guide entrance. These findings align with our results. Furthermore, a slightly larger moderator and guide cross-section help mitigate flux losses caused by guide misalignments. Overall, comparing Pioneer's results with the COFSI function underscores the effectiveness of our optimization process.

A neutron transport system aims at maximizing the delivery of neutrons within the required phase space to enhance signal while minimizing undesired neutrons outside this phase space as early as possible. This prioritizes transport efficiency and reduces downstream shielding requirements. Consequently, starting the guide as close as possible to the moderator to maximize the acceptance angle is often unnecessary. Feeding excessive neutrons into the guide increases shielding requirement, as neutrons outside the required phase space must be filtered out. Additionally, positioning the guide farther from the moderator helps mitigate radiation damage to the guide substrates~\cite{rowe2005analysis}. For Pioneer, the STS core vessel allows the instrument optics to begin at 1.02~m from the moderator. However, our optimization indicates that the guide can start at 2.59~m while still extracting sufficient neutrons. Therefore, we chose to include a pre-guide steel collimator to reduce the undesired neutrons impinging on the guide substrates or entering the guide system~\cite{liu2025optical}. At the same time, our optimization demonstrates that fine-tuning of the guide geometry can effectively mitigate the effects of in-guide gaps that can potentially introduce divergence discontinuities.

To further quantify the performance of the optmized guide, we have calculated the brilliance transfer (BT), shown in Fig.~\ref{fig:BT}. BT quantifies the neutron transport efficiency of the optical system from the source to the sample. Based on Liouville's theorem, the theoretical maximum BT is 100\% for a perfect and conservative beam transport system~\cite{andersen2018optimization}. Imperfections within the system can significantly reduce this value. Pioneer exhibits high BT values of 93-95\%, depending on the chosen region of interest. The slight reduction from the ideal 100\% is primarily due to the imperfect guide reflectivity used in the simulations~\cite{liu2024general}, which assumed at least 1\% beam loss per bounce. These high BT values confirm the high effectiveness of the optimized guide geometry for neutron transport, even with the presence of large in-guide gaps.

A high BT value correlates with high phase space homogeneity. Assuming uniform source brilliance, phase space inhomogeneity necessarily leads to lower average brilliance transfer (due to Liouville's theorem). This effect is well illustrated from the comparison of straight-elliptical and curved-tapered beamlines in our previous work~\cite{liu2024general}. Furthermore, spatial and angular homogeneity are important for many single crystal diffraction experiments. Spatial homogeneity ensures the applicability of calibration spectra across all sample sizes and minor misalignments, while angular homogeneity reduces ambiguity in scattering features~\cite{liu2024general}.

Many modern ballistic guide designs incorporate curved profiles. While truly curved guides can be engineered at a higher cost, they can be approximated using polygonal segments composed of straight supermirror elements. For Pioneer, our optimization revealed that using straight supermirror elements approximately $0.5$~m in length provides satisfactory neutron transport efficiency and beam homogeneity. Only a few shorter elements are needed to accommodate constraints related to the STS monolith and bunker walls. 

The choice of the supermirror $m$-value significantly influences neutron transport performance. Supermirrors with $m$-values up to 8 are commercially available~\cite{swissneutronics}, and high-$m$ supermirrors are crucial for transporting high-divergence beams and short-wavelength neutrons. However, high-$m$ supermirrors come with notable drawbacks. They are substantially more expensive and present two additional disadvantages. First, higher-$m$ supermirrors exhibit a steeper decline in reflectivity in the low-$Q$ region~\cite{jacobsen2013bi, swissneutronics}, leading to greater flux loss on reflection and an increased number of leaked neutrons requiring additional shielding. Second, experimental studies have demonstrated that higher-$m$ supermirrors cause increased neutron absorption within the multilayers~\cite{kolevatov2019neutron}, resulting in elevated levels of high-energy gamma rays that also necessitate more shielding. Overall, to achieve optimal neutron transport performance, the lowest $m$-value that fulfills the phase space requirements is the optimal choice.

By comparing the simulated performance of the polarizing V-cavity and $^3$He spin filters, we found that the polarizing V-cavity option offers a higher average FOM than $^3$He spin filters over 1.2-5.5~\AA, even when considering a favorable scenario for $^3$He spin filters of a high $^3$He polarization. The polarizing V-cavity maintains a high and gradually varying FOM across the entire band, whereas the $^3$He spin filter exhibits a stronger wavelength dependence. The optimal $p_\text{He} \cdot l$ value depends on the specific choice of $FOM_\text{pol, eff}$, but it does not alter the overall conclusion.  Therefore, Pioneer will use a polarizing V-cavity as the default polarizer. 

However, polarizing V-cavities have a cutoff wavelength below which their polarization capability is lost. In contrast, $^3$He spin filters do not exhibit such a cutoff, offering flexibility to provide polarized beams at shorter wavelengths. To ensure operational adaptability, two gate valves isolate the polarizer vessel from the guide~\cite{liu2025optical}. This design allows the vessel to be removed from the beamline, leaving space to mount a $^3$He spin filter as an upgrade path to extend the polarized beam range down to 0.8~\AA.

In summary, Pioneer will be a next-generation single-crystal neutron diffractometer to study small-volume samples and weak signals in versatile sample environments. We have conducted Montel Carlo ray tracing simulations to optimize the neutron transport and control system for the incident beam, aiming to deliver high flux, homogeneous beams with tunable beam size and divergence to address a wide range scientific needs. For the incident beam polarizer, a three-channel polarizing V-cavity will provide a high performance, with an average $P^2T \sim 35\%$ over the default polarized band of 1.2-5.5~\AA. 

\section{Supplementary Materials}
The supplementary materials include 3 figures. Figure~\ref{fig:BT} displays the source brilliance and the brilliance transfers calculated in selected regions of interest at the sample position. Figure~\ref{fig:VE_QRes} shows the results of a virtual experiment in the HR mode. Figure~\ref{fig:PolRef} plots the spin dependent reflectivity of the polarizing supermirrors used in Monte Carlo ray tracing simulations.

\begin{acknowledgments}
Liu thanks Stephan Rosenkranz, along with the other members of the instrument advisory committee for their discussions, and acknowledges Fankang Li and Peter Jiang for recommendations about the polarization systems. The authors also thank Kenneth Herwig, Leighton Coates, Scott Dixon, and Van Graves for valuable comments on the instrument design. This research used resources of the Spallation Neutron Source Second Target Station Project at Oak Ridge National Laboratory (ORNL). ORNL is managed by UT-Battelle LLC for DOE's Office of Science, the single largest supporter of basic research in the physical sciences in the United States. This manuscript has been authored by UT-Battelle, LLC, under contract DE-AC05-00OR22725 with the US Department of Energy (DOE; Basic Energy Sciences).
\end{acknowledgments}

\section*{Data Availability Statement}
The data that support the findings of this study are available from the corresponding author upon reasonable request.

\pagebreak

\renewcommand{\thefigure}{S\arabic{figure}}
\setcounter{figure}{0}
\begin{CJK*}{GB}{} 

\title[]{Supplementary Materials: Incident beam optics optimization for the single crystal neutron diffractometer Pioneer with a polarized beam option}

\author{Yaohua Liu*}
\email[Author to whom correspondence should be addressed: ]{liuyh@ornl.gov}
\affiliation{Second Target Station, Oak Ridge National Laboratory, Oak Ridge, Tennessee 37831, USA}%

\author{Peter Torres}
\affiliation{Second Target Station, Oak Ridge National Laboratory, Oak Ridge, Tennessee 37831, USA}%

\date{\today}
\maketitle
\end{CJK*}

\begin{figure*}
\includegraphics[width=1.0\textwidth]{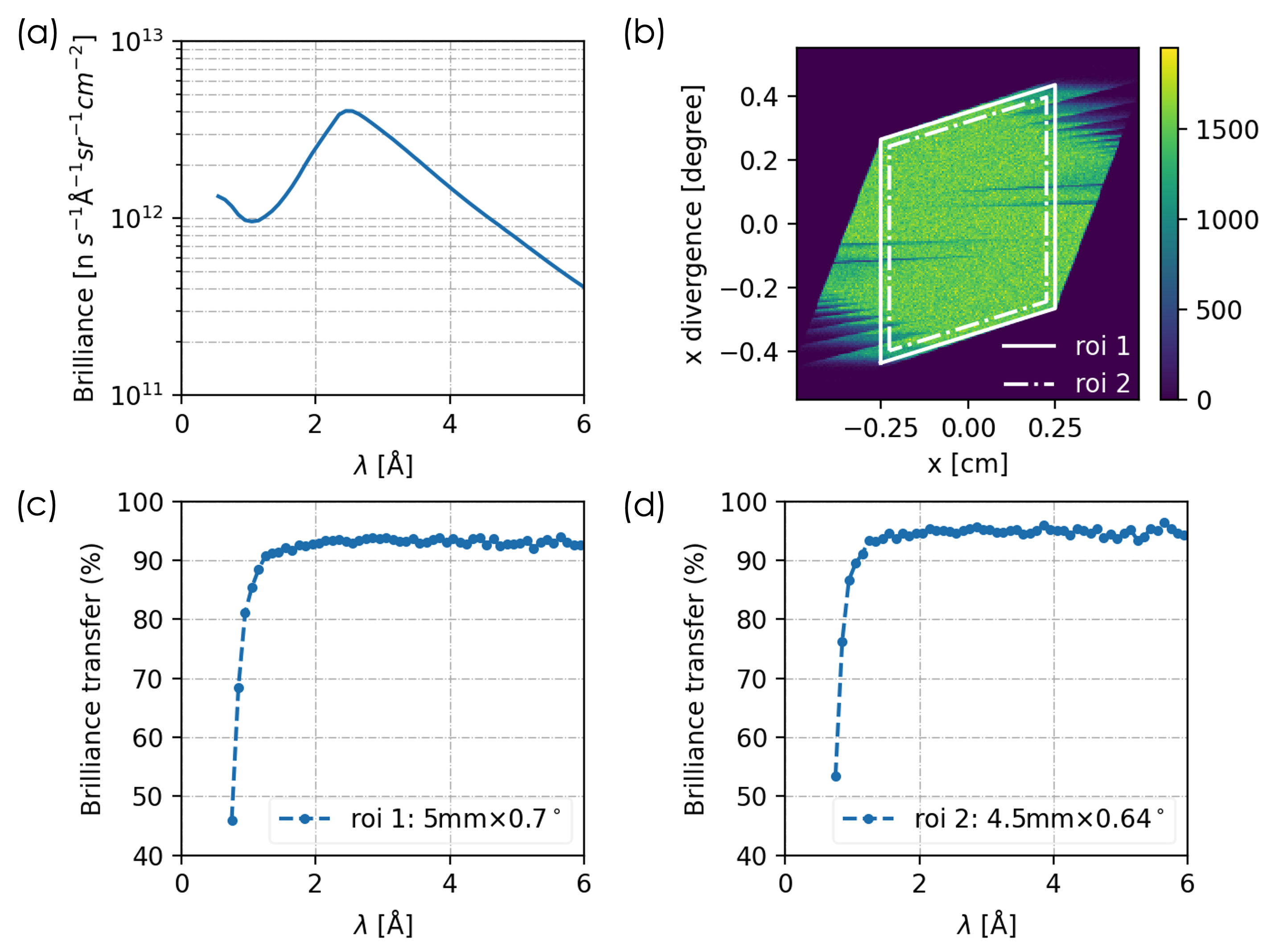}
\caption{\label{fig:BT}(a) Simulated STS source brilliance from the cylindrical moderator using the source file \textit{BL13-CY-46D-STS-Min-2G-source\_mctal-125\_sp.dat}. (b) Horizontal (x-direction) acceptance diagram at the sample position.  The solid and dot-dashed lines denote two regions of interest (ROIs) used for brilliance transfer estimation. The solid box (ROI 1) shows the full phase space volume required by the HF-LG sciences, i.e., $0.70^\circ \times 5.0$~mm in both the x and y directions. The dot-dashed box (ROI 2) represents a reduced phase space ($0.64^\circ \times 4.5$~mm in both the x and y directions) chosen to avoid the edge effect. (c) and (d) Brilliance transfers with saturated values of 93\% and 95\% evaluated over ROI 1 and ROI 2, respectively.}
\end{figure*}

\begin{figure*}
\includegraphics[width=1.0\textwidth]{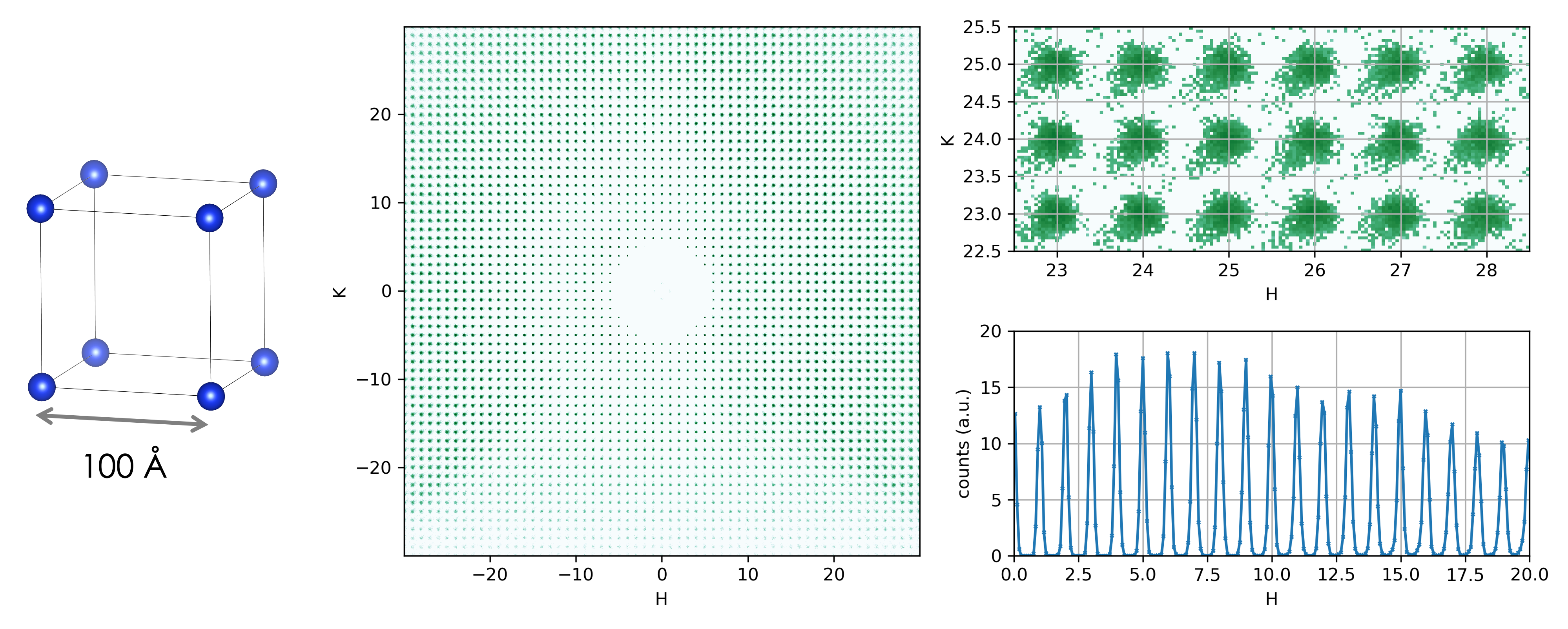}
\caption{\label{fig:VE_QRes} Virtual single-crystal diffraction experiment simulating a simple cubic monatomic crystal with a lattice constant of 100~\AA~(Left) in the HR mode. The crystal has a mosaic spread of 0.3$^\circ$ and dimensions of $0.5 \times 0.5 \times 0.5$~mm$^3$. (Middle) and (Top Right): Simulated diffraction patterns covering a large and a small region of reciprocal space, respectively. (Bottom Right): Line cut illustrating well-separated Bragg peaks.}
\end{figure*}

\begin{figure*}
\includegraphics[width=0.8\textwidth]{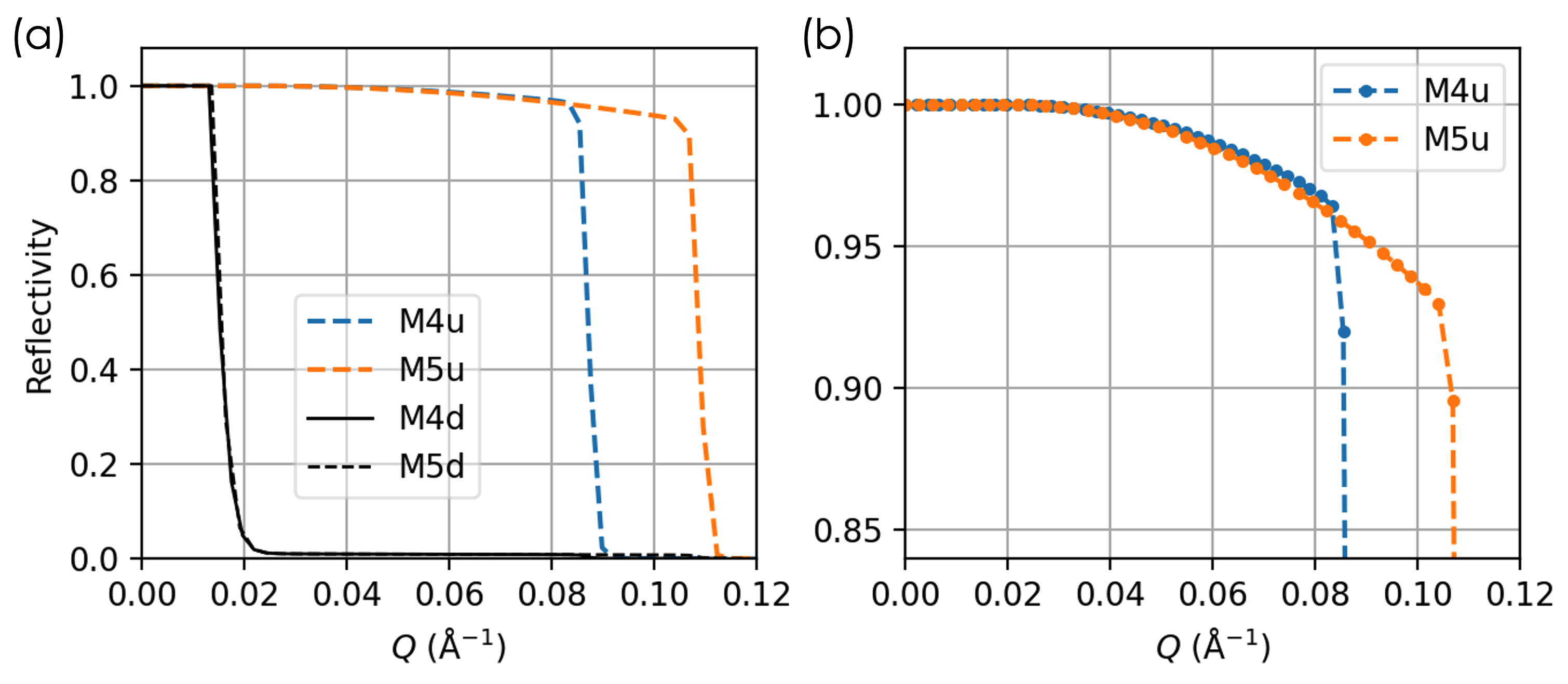}
\caption{\label{fig:PolRef} Spin dependent reflectivity for double-side coated $m=4, 5$ Fe/Si polarizing supermirrors. "u" and "d" stand for up and down spin states, respectively.}
\end{figure*}


\begin{thebibliography}{36}%
\makeatletter
\providecommand \@ifxundefined [1]{%
 \@ifx{#1\undefined}
}%
\providecommand \@ifnum [1]{%
 \ifnum #1\expandafter \@firstoftwo
 \else \expandafter \@secondoftwo
 \fi
}%
\providecommand \@ifx [1]{%
 \ifx #1\expandafter \@firstoftwo
 \else \expandafter \@secondoftwo
 \fi
}%
\providecommand \natexlab [1]{#1}%
\providecommand \enquote  [1]{``#1''}%
\providecommand \bibnamefont  [1]{#1}%
\providecommand \bibfnamefont [1]{#1}%
\providecommand \citenamefont [1]{#1}%
\providecommand \href@noop [0]{\@secondoftwo}%
\providecommand \href [0]{\begingroup \@sanitize@url \@href}%
\providecommand \@href[1]{\@@startlink{#1}\@@href}%
\providecommand \@@href[1]{\endgroup#1\@@endlink}%
\providecommand \@sanitize@url [0]{\catcode `\\12\catcode `\$12\catcode
  `\&12\catcode `\#12\catcode `\^12\catcode `\_12\catcode `\%12\relax}%
\providecommand \@@startlink[1]{}%
\providecommand \@@endlink[0]{}%
\providecommand \url  [0]{\begingroup\@sanitize@url \@url }%
\providecommand \@url [1]{\endgroup\@href {#1}{\urlprefix }}%
\providecommand \urlprefix  [0]{URL }%
\providecommand \Eprint [0]{\href }%
\providecommand \doibase [0]{http://dx.doi.org/}%
\providecommand \selectlanguage [0]{\@gobble}%
\providecommand \bibinfo  [0]{\@secondoftwo}%
\providecommand \bibfield  [0]{\@secondoftwo}%
\providecommand \translation [1]{[#1]}%
\providecommand \BibitemOpen [0]{}%
\providecommand \bibitemStop [0]{}%
\providecommand \bibitemNoStop [0]{.\EOS\space}%
\providecommand \EOS [0]{\spacefactor3000\relax}%
\providecommand \BibitemShut  [1]{\csname bibitem#1\endcsname}%
\let\auto@bib@innerbib\@empty
\bibitem [{\citenamefont {ORNL}(2020)}]{adams2020first}%
  \BibitemOpen
  \bibfield  {author} {\bibinfo {author} {\bibnamefont {ORNL}},\ }\href
  {https://neutrons.ornl.gov/sites/default/files/STS_First_Experiments_Report.pdf}
  {\emph {\bibinfo {title} {{First Experiments: New Science Opportunities at
  the Spallation Neutron Source Second Target Station (abridged)}}}},\ \bibinfo
  {type} {Tech. Rep.}\ (\bibinfo  {institution} {Oak Ridge National Lab., Oak
  Ridge, TN (United States)},\ \bibinfo {year} {2020})\BibitemShut {NoStop}%
\bibitem [{\citenamefont {Wilson}(2000)}]{wilson2000single}%
  \BibitemOpen
  \bibfield  {author} {\bibinfo {author} {\bibfnamefont {C.~C.}\ \bibnamefont
  {Wilson}},\ }\href@noop {} {\emph {\bibinfo {title} {Single crystal neutron
  diffraction from molecular materials}}},\ Vol.~\bibinfo {volume} {2}\
  (\bibinfo  {publisher} {World Scientific},\ \bibinfo {year}
  {2000})\BibitemShut {NoStop}%
\bibitem [{\citenamefont {Nield}\ and\ \citenamefont
  {Keen}(2001)}]{nield2001diffuse}%
  \BibitemOpen
  \bibfield  {author} {\bibinfo {author} {\bibfnamefont {V.~M.}\ \bibnamefont
  {Nield}}\ and\ \bibinfo {author} {\bibfnamefont {D.~A.}\ \bibnamefont
  {Keen}},\ }\href@noop {} {\emph {\bibinfo {title} {Diffuse neutron scattering
  from crystalline materials}}},\ Vol.~\bibinfo {volume} {14}\ (\bibinfo
  {publisher} {Oxford University Press},\ \bibinfo {year} {2001})\BibitemShut
  {NoStop}%
\bibitem [{\citenamefont {Carpenter}\ and\ \citenamefont
  {Loong}(2015)}]{carpenter2015elements}%
  \BibitemOpen
  \bibfield  {author} {\bibinfo {author} {\bibfnamefont {J.~M.}\ \bibnamefont
  {Carpenter}}\ and\ \bibinfo {author} {\bibfnamefont {C.-K.}\ \bibnamefont
  {Loong}},\ }\href@noop {} {\emph {\bibinfo {title} {Elements of slow-neutron
  scattering}}}\ (\bibinfo  {publisher} {Cambridge University Press},\ \bibinfo
  {year} {2015})\BibitemShut {NoStop}%
\bibitem [{\citenamefont {Liu}\ \emph {et~al.}(2022)\citenamefont {Liu},
  \citenamefont {Cao}, \citenamefont {Rosenkranz}, \citenamefont {Frost},
  \citenamefont {Huegle}, \citenamefont {Lin}, \citenamefont {Torres},
  \citenamefont {Stoica},\ and\ \citenamefont {Chakoumakos}}]{liu2022Pioneer}%
  \BibitemOpen
  \bibfield  {author} {\bibinfo {author} {\bibfnamefont {Y.}~\bibnamefont
  {Liu}}, \bibinfo {author} {\bibfnamefont {H.}~\bibnamefont {Cao}}, \bibinfo
  {author} {\bibfnamefont {S.}~\bibnamefont {Rosenkranz}}, \bibinfo {author}
  {\bibfnamefont {M.}~\bibnamefont {Frost}}, \bibinfo {author} {\bibfnamefont
  {T.}~\bibnamefont {Huegle}}, \bibinfo {author} {\bibfnamefont {J.~Y.~Y.}\
  \bibnamefont {Lin}}, \bibinfo {author} {\bibfnamefont {P.}~\bibnamefont
  {Torres}}, \bibinfo {author} {\bibfnamefont {A.}~\bibnamefont {Stoica}}, \
  and\ \bibinfo {author} {\bibfnamefont {B.~C.}\ \bibnamefont {Chakoumakos}},\
  }\href {\doibase 10.1063/5.0089524} {\bibfield  {journal} {\bibinfo
  {journal} {Review of Scientific Instruments}\ }\textbf {\bibinfo {volume}
  {93}},\ \bibinfo {pages} {073901} (\bibinfo {year} {2022})}\BibitemShut
  {NoStop}%
\bibitem [{\citenamefont {Konik}\ and\ \citenamefont
  {Ioffe}(2023)}]{konik2023new}%
  \BibitemOpen
  \bibfield  {author} {\bibinfo {author} {\bibfnamefont {P.}~\bibnamefont
  {Konik}}\ and\ \bibinfo {author} {\bibfnamefont {A.}~\bibnamefont {Ioffe}},\
  }\href@noop {} {\bibfield  {journal} {\bibinfo  {journal} {Nuclear
  Instruments and Methods in Physics Research Section A: Accelerators,
  Spectrometers, Detectors and Associated Equipment}\ }\textbf {\bibinfo
  {volume} {1056}},\ \bibinfo {pages} {168643} (\bibinfo {year}
  {2023})}\BibitemShut {NoStop}%
\bibitem [{\citenamefont {Walters}\ \emph {et~al.}(2009)\citenamefont
  {Walters}, \citenamefont {Perring}, \citenamefont {Caux}, \citenamefont
  {Savici}, \citenamefont {Gu}, \citenamefont {Lee}, \citenamefont {Ku},\ and\
  \citenamefont {Zaliznyak}}]{walters2009effect}%
  \BibitemOpen
  \bibfield  {author} {\bibinfo {author} {\bibfnamefont {A.~C.}\ \bibnamefont
  {Walters}}, \bibinfo {author} {\bibfnamefont {T.~G.}\ \bibnamefont
  {Perring}}, \bibinfo {author} {\bibfnamefont {J.-S.}\ \bibnamefont {Caux}},
  \bibinfo {author} {\bibfnamefont {A.~T.}\ \bibnamefont {Savici}}, \bibinfo
  {author} {\bibfnamefont {G.~D.}\ \bibnamefont {Gu}}, \bibinfo {author}
  {\bibfnamefont {C.-C.}\ \bibnamefont {Lee}}, \bibinfo {author} {\bibfnamefont
  {W.}~\bibnamefont {Ku}}, \ and\ \bibinfo {author} {\bibfnamefont {I.~A.}\
  \bibnamefont {Zaliznyak}},\ }\href@noop {} {\bibfield  {journal} {\bibinfo
  {journal} {Nature Physics}\ }\textbf {\bibinfo {volume} {5}},\ \bibinfo
  {pages} {867} (\bibinfo {year} {2009})}\BibitemShut {NoStop}%
\bibitem [{\citenamefont {Liu}\ \emph {et~al.}(2025)\citenamefont {Liu},
  \citenamefont {Torres}, \citenamefont {Dixon}, \citenamefont {Darian},
  \citenamefont {Khaplanov}, \citenamefont {Mchargue},\ and\ \citenamefont
  {Thermer}}]{liu2025optical}%
  \BibitemOpen
  \bibfield  {author} {\bibinfo {author} {\bibfnamefont {Y.}~\bibnamefont
  {Liu}}, \bibinfo {author} {\bibfnamefont {P.}~\bibnamefont {Torres}},
  \bibinfo {author} {\bibfnamefont {S.}~\bibnamefont {Dixon}}, \bibinfo
  {author} {\bibfnamefont {K.}~\bibnamefont {Darian}}, \bibinfo {author}
  {\bibfnamefont {A.}~\bibnamefont {Khaplanov}}, \bibinfo {author}
  {\bibfnamefont {B.}~\bibnamefont {Mchargue}}, \ and\ \bibinfo {author}
  {\bibfnamefont {R.}~\bibnamefont {Thermer}},\ }\href {\doibase
  10.1063/5.0247408} {\bibfield  {journal} {\bibinfo  {journal} {Review of
  Scientific Instruments}\ }\textbf {\bibinfo {volume} {96}},\ \bibinfo {pages}
  {in press} (\bibinfo {year} {2025})}\BibitemShut {NoStop}%
\bibitem [{\citenamefont {Liu}(2024)}]{liu2024general}%
  \BibitemOpen
  \bibfield  {author} {\bibinfo {author} {\bibfnamefont {Y.}~\bibnamefont
  {Liu}},\ }\href {\doibase 10.1063/5.0212920} {\bibfield  {journal} {\bibinfo
  {journal} {Review of Scientific Instruments}\ }\textbf {\bibinfo {volume}
  {95}},\ \bibinfo {pages} {073902} (\bibinfo {year} {2024})}\BibitemShut
  {NoStop}%
\bibitem [{\citenamefont {Mezei}(1976)}]{mezei1976novel}%
  \BibitemOpen
  \bibfield  {author} {\bibinfo {author} {\bibfnamefont {F.}~\bibnamefont
  {Mezei}},\ }\href@noop {} {\bibfield  {journal} {\bibinfo  {journal}
  {Communications on Physics (London)}\ }\textbf {\bibinfo {volume} {1}},\
  \bibinfo {pages} {81} (\bibinfo {year} {1976})}\BibitemShut {NoStop}%
\bibitem [{\citenamefont {Willendrup}\ \emph {et~al.}(2014)\citenamefont
  {Willendrup}, \citenamefont {Farhi}, \citenamefont {Knudsen}, \citenamefont
  {Filges},\ and\ \citenamefont {Lefmann}}]{willendrup2014mcstas}%
  \BibitemOpen
  \bibfield  {author} {\bibinfo {author} {\bibfnamefont {P.}~\bibnamefont
  {Willendrup}}, \bibinfo {author} {\bibfnamefont {E.}~\bibnamefont {Farhi}},
  \bibinfo {author} {\bibfnamefont {E.}~\bibnamefont {Knudsen}}, \bibinfo
  {author} {\bibfnamefont {U.}~\bibnamefont {Filges}}, \ and\ \bibinfo {author}
  {\bibfnamefont {K.}~\bibnamefont {Lefmann}},\ }\href@noop {} {\bibfield
  {journal} {\bibinfo  {journal} {Journal of Neutron Research}\ }\textbf
  {\bibinfo {volume} {17}},\ \bibinfo {pages} {35} (\bibinfo {year}
  {2014})}\BibitemShut {NoStop}%
\bibitem [{\citenamefont {Lieutenant}\ \emph {et~al.}(2004)\citenamefont
  {Lieutenant}, \citenamefont {Zsigmond}, \citenamefont {Manoshin},
  \citenamefont {Fromme}, \citenamefont {Bordallo}, \citenamefont {Champion},
  \citenamefont {Peters},\ and\ \citenamefont {Mezei}}]{lieutenant2004neutron}%
  \BibitemOpen
  \bibfield  {author} {\bibinfo {author} {\bibfnamefont {K.}~\bibnamefont
  {Lieutenant}}, \bibinfo {author} {\bibfnamefont {G.}~\bibnamefont
  {Zsigmond}}, \bibinfo {author} {\bibfnamefont {S.}~\bibnamefont {Manoshin}},
  \bibinfo {author} {\bibfnamefont {M.}~\bibnamefont {Fromme}}, \bibinfo
  {author} {\bibfnamefont {H.~N.}\ \bibnamefont {Bordallo}}, \bibinfo {author}
  {\bibfnamefont {D.}~\bibnamefont {Champion}}, \bibinfo {author}
  {\bibfnamefont {J.}~\bibnamefont {Peters}}, \ and\ \bibinfo {author}
  {\bibfnamefont {F.}~\bibnamefont {Mezei}},\ }in\ \href@noop {} {\emph
  {\bibinfo {booktitle} {Advances in Computational Methods for X-Ray and
  Neutron Optics}}},\ Vol.\ \bibinfo {volume} {5536}\ (\bibinfo {organization}
  {SPIE},\ \bibinfo {year} {2004})\ pp.\ \bibinfo {pages}
  {134--145}\BibitemShut {NoStop}%
\bibitem [{\citenamefont {Lin}\ \emph {et~al.}(2016)\citenamefont {Lin},
  \citenamefont {Smith}, \citenamefont {Granroth}, \citenamefont {Abernathy},
  \citenamefont {Lumsden}, \citenamefont {Winn}, \citenamefont {Aczel},
  \citenamefont {Aivazis},\ and\ \citenamefont {Fultz}}]{lin2016MCViNE}%
  \BibitemOpen
  \bibfield  {author} {\bibinfo {author} {\bibfnamefont {J.~Y.}\ \bibnamefont
  {Lin}}, \bibinfo {author} {\bibfnamefont {H.~L.}\ \bibnamefont {Smith}},
  \bibinfo {author} {\bibfnamefont {G.~E.}\ \bibnamefont {Granroth}}, \bibinfo
  {author} {\bibfnamefont {D.~L.}\ \bibnamefont {Abernathy}}, \bibinfo {author}
  {\bibfnamefont {M.~D.}\ \bibnamefont {Lumsden}}, \bibinfo {author}
  {\bibfnamefont {B.}~\bibnamefont {Winn}}, \bibinfo {author} {\bibfnamefont
  {A.~A.}\ \bibnamefont {Aczel}}, \bibinfo {author} {\bibfnamefont
  {M.}~\bibnamefont {Aivazis}}, \ and\ \bibinfo {author} {\bibfnamefont
  {B.}~\bibnamefont {Fultz}},\ }\href@noop {} {\bibfield  {journal} {\bibinfo
  {journal} {Nuclear Instruments and Methods in Physics Research Section A:
  Accelerators, Spectrometers, Detectors and Associated Equipment}\ }\textbf
  {\bibinfo {volume} {810}},\ \bibinfo {pages} {86} (\bibinfo {year}
  {2016})}\BibitemShut {NoStop}%
\bibitem [{\citenamefont {Hayter}\ and\ \citenamefont
  {Mook}(1989)}]{hayter1989discrete}%
  \BibitemOpen
  \bibfield  {author} {\bibinfo {author} {\bibfnamefont {J.~B.}\ \bibnamefont
  {Hayter}}\ and\ \bibinfo {author} {\bibfnamefont {H.}~\bibnamefont {Mook}},\
  }\href@noop {} {\bibfield  {journal} {\bibinfo  {journal} {Journal of Applied
  Crystallography}\ }\textbf {\bibinfo {volume} {22}},\ \bibinfo {pages} {35}
  (\bibinfo {year} {1989})}\BibitemShut {NoStop}%
\bibitem [{\citenamefont {Schebetov}\ \emph {et~al.}(1999)\citenamefont
  {Schebetov}, \citenamefont {Kovalev}, \citenamefont {Peskov}, \citenamefont
  {Pleshanov}, \citenamefont {Pusenkov}, \citenamefont {Schubert-Bischoff},
  \citenamefont {Shmelev}, \citenamefont {Soroko}, \citenamefont
  {Syromyatnikov}, \citenamefont {Ul'yanov} \emph
  {et~al.}}]{schebetov1999multi}%
  \BibitemOpen
  \bibfield  {author} {\bibinfo {author} {\bibfnamefont {A.}~\bibnamefont
  {Schebetov}}, \bibinfo {author} {\bibfnamefont {A.}~\bibnamefont {Kovalev}},
  \bibinfo {author} {\bibfnamefont {B.}~\bibnamefont {Peskov}}, \bibinfo
  {author} {\bibfnamefont {N.}~\bibnamefont {Pleshanov}}, \bibinfo {author}
  {\bibfnamefont {V.}~\bibnamefont {Pusenkov}}, \bibinfo {author}
  {\bibfnamefont {P.}~\bibnamefont {Schubert-Bischoff}}, \bibinfo {author}
  {\bibfnamefont {G.}~\bibnamefont {Shmelev}}, \bibinfo {author} {\bibfnamefont
  {Z.}~\bibnamefont {Soroko}}, \bibinfo {author} {\bibfnamefont
  {V.}~\bibnamefont {Syromyatnikov}}, \bibinfo {author} {\bibfnamefont
  {V.}~\bibnamefont {Ul'yanov}},  \emph {et~al.},\ }\href@noop {} {\bibfield
  {journal} {\bibinfo  {journal} {Nuclear Instruments and Methods in Physics
  Research Section A: Accelerators, Spectrometers, Detectors and Associated
  Equipment}\ }\textbf {\bibinfo {volume} {432}},\ \bibinfo {pages} {214}
  (\bibinfo {year} {1999})}\BibitemShut {NoStop}%
\bibitem [{\citenamefont {Padiyath}\ \emph {et~al.}(2004)\citenamefont
  {Padiyath}, \citenamefont {Stahn}, \citenamefont {Allenspach}, \citenamefont
  {Horisberger},\ and\ \citenamefont {B{\"o}ni}}]{padiyath2004influence}%
  \BibitemOpen
  \bibfield  {author} {\bibinfo {author} {\bibfnamefont {J.}~\bibnamefont
  {Padiyath}}, \bibinfo {author} {\bibfnamefont {J.}~\bibnamefont {Stahn}},
  \bibinfo {author} {\bibfnamefont {P.}~\bibnamefont {Allenspach}}, \bibinfo
  {author} {\bibfnamefont {M.}~\bibnamefont {Horisberger}}, \ and\ \bibinfo
  {author} {\bibfnamefont {P.}~\bibnamefont {B{\"o}ni}},\ }\href@noop {}
  {\bibfield  {journal} {\bibinfo  {journal} {Physica B: Condensed Matter}\
  }\textbf {\bibinfo {volume} {350}},\ \bibinfo {pages} {E237} (\bibinfo {year}
  {2004})}\BibitemShut {NoStop}%
\bibitem [{\citenamefont {Kovacs-Mezei}\ \emph {et~al.}(2008)\citenamefont
  {Kovacs-Mezei}, \citenamefont {Krist},\ and\ \citenamefont
  {R{\'e}vay}}]{kovacs2008non}%
  \BibitemOpen
  \bibfield  {author} {\bibinfo {author} {\bibfnamefont {R.}~\bibnamefont
  {Kovacs-Mezei}}, \bibinfo {author} {\bibfnamefont {T.}~\bibnamefont {Krist}},
  \ and\ \bibinfo {author} {\bibfnamefont {Z.}~\bibnamefont {R{\'e}vay}},\
  }\href@noop {} {\bibfield  {journal} {\bibinfo  {journal} {Nuclear
  Instruments and Methods in Physics Research Section A: Accelerators,
  Spectrometers, Detectors and Associated Equipment}\ }\textbf {\bibinfo
  {volume} {586}},\ \bibinfo {pages} {51} (\bibinfo {year} {2008})}\BibitemShut
  {NoStop}%
\bibitem [{\citenamefont {Schanzer}\ \emph {et~al.}(2016)\citenamefont
  {Schanzer}, \citenamefont {Schneider},\ and\ \citenamefont
  {B{\"o}ni}}]{schanzer2016neutron}%
  \BibitemOpen
  \bibfield  {author} {\bibinfo {author} {\bibfnamefont {C.}~\bibnamefont
  {Schanzer}}, \bibinfo {author} {\bibfnamefont {M.}~\bibnamefont {Schneider}},
  \ and\ \bibinfo {author} {\bibfnamefont {P.}~\bibnamefont {B{\"o}ni}},\ }in\
  \href@noop {} {\emph {\bibinfo {booktitle} {Journal of Physics: Conference
  Series}}},\ Vol.\ \bibinfo {volume} {746}\ (\bibinfo {organization} {IOP
  Publishing},\ \bibinfo {year} {2016})\ p.\ \bibinfo {pages}
  {012024}\BibitemShut {NoStop}%
\bibitem [{\citenamefont {Virtanen}\ \emph {et~al.}(2020)\citenamefont
  {Virtanen}, \citenamefont {Gommers}, \citenamefont {Oliphant}, \citenamefont
  {Haberland}, \citenamefont {Reddy}, \citenamefont {Cournapeau}, \citenamefont
  {Burovski}, \citenamefont {Peterson}, \citenamefont {Weckesser},
  \citenamefont {Bright}, \citenamefont {{van der Walt}}, \citenamefont
  {Brett}, \citenamefont {Wilson}, \citenamefont {Millman}, \citenamefont
  {Mayorov}, \citenamefont {Nelson}, \citenamefont {Jones}, \citenamefont
  {Kern}, \citenamefont {Larson}, \citenamefont {Carey}, \citenamefont {Polat},
  \citenamefont {Feng}, \citenamefont {Moore}, \citenamefont {{VanderPlas}},
  \citenamefont {Laxalde}, \citenamefont {Perktold}, \citenamefont {Cimrman},
  \citenamefont {Henriksen}, \citenamefont {Quintero}, \citenamefont {Harris},
  \citenamefont {Archibald}, \citenamefont {Ribeiro}, \citenamefont
  {Pedregosa}, \citenamefont {{van Mulbregt}},\ and\ \citenamefont {{SciPy 1.0
  Contributors}}}]{2020SciPy-NMeth}%
  \BibitemOpen
  \bibfield  {author} {\bibinfo {author} {\bibfnamefont {P.}~\bibnamefont
  {Virtanen}}, \bibinfo {author} {\bibfnamefont {R.}~\bibnamefont {Gommers}},
  \bibinfo {author} {\bibfnamefont {T.~E.}\ \bibnamefont {Oliphant}}, \bibinfo
  {author} {\bibfnamefont {M.}~\bibnamefont {Haberland}}, \bibinfo {author}
  {\bibfnamefont {T.}~\bibnamefont {Reddy}}, \bibinfo {author} {\bibfnamefont
  {D.}~\bibnamefont {Cournapeau}}, \bibinfo {author} {\bibfnamefont
  {E.}~\bibnamefont {Burovski}}, \bibinfo {author} {\bibfnamefont
  {P.}~\bibnamefont {Peterson}}, \bibinfo {author} {\bibfnamefont
  {W.}~\bibnamefont {Weckesser}}, \bibinfo {author} {\bibfnamefont
  {J.}~\bibnamefont {Bright}}, \bibinfo {author} {\bibfnamefont {S.~J.}\
  \bibnamefont {{van der Walt}}}, \bibinfo {author} {\bibfnamefont
  {M.}~\bibnamefont {Brett}}, \bibinfo {author} {\bibfnamefont
  {J.}~\bibnamefont {Wilson}}, \bibinfo {author} {\bibfnamefont {K.~J.}\
  \bibnamefont {Millman}}, \bibinfo {author} {\bibfnamefont {N.}~\bibnamefont
  {Mayorov}}, \bibinfo {author} {\bibfnamefont {A.~R.~J.}\ \bibnamefont
  {Nelson}}, \bibinfo {author} {\bibfnamefont {E.}~\bibnamefont {Jones}},
  \bibinfo {author} {\bibfnamefont {R.}~\bibnamefont {Kern}}, \bibinfo {author}
  {\bibfnamefont {E.}~\bibnamefont {Larson}}, \bibinfo {author} {\bibfnamefont
  {C.~J.}\ \bibnamefont {Carey}}, \bibinfo {author} {\bibfnamefont
  {{\.I}.}~\bibnamefont {Polat}}, \bibinfo {author} {\bibfnamefont
  {Y.}~\bibnamefont {Feng}}, \bibinfo {author} {\bibfnamefont {E.~W.}\
  \bibnamefont {Moore}}, \bibinfo {author} {\bibfnamefont {J.}~\bibnamefont
  {{VanderPlas}}}, \bibinfo {author} {\bibfnamefont {D.}~\bibnamefont
  {Laxalde}}, \bibinfo {author} {\bibfnamefont {J.}~\bibnamefont {Perktold}},
  \bibinfo {author} {\bibfnamefont {R.}~\bibnamefont {Cimrman}}, \bibinfo
  {author} {\bibfnamefont {I.}~\bibnamefont {Henriksen}}, \bibinfo {author}
  {\bibfnamefont {E.~A.}\ \bibnamefont {Quintero}}, \bibinfo {author}
  {\bibfnamefont {C.~R.}\ \bibnamefont {Harris}}, \bibinfo {author}
  {\bibfnamefont {A.~M.}\ \bibnamefont {Archibald}}, \bibinfo {author}
  {\bibfnamefont {A.~H.}\ \bibnamefont {Ribeiro}}, \bibinfo {author}
  {\bibfnamefont {F.}~\bibnamefont {Pedregosa}}, \bibinfo {author}
  {\bibfnamefont {P.}~\bibnamefont {{van Mulbregt}}}, \ and\ \bibinfo {author}
  {\bibnamefont {{SciPy 1.0 Contributors}}},\ }\href {\doibase
  10.1038/s41592-019-0686-2} {\bibfield  {journal} {\bibinfo  {journal} {Nature
  Methods}\ }\textbf {\bibinfo {volume} {17}},\ \bibinfo {pages} {261}
  (\bibinfo {year} {2020})}\BibitemShut {NoStop}%
\bibitem [{\citenamefont {Kluyver}\ \emph {et~al.}(2016)\citenamefont
  {Kluyver}, \citenamefont {Ragan-Kelley}, \citenamefont {P{\'e}rez},
  \citenamefont {Granger}, \citenamefont {Bussonnier}, \citenamefont
  {Frederic}, \citenamefont {Kelley}, \citenamefont {Hamrick}, \citenamefont
  {Grout}, \citenamefont {Corlay} \emph {et~al.}}]{kluyver2016jupyter}%
  \BibitemOpen
  \bibfield  {author} {\bibinfo {author} {\bibfnamefont {T.}~\bibnamefont
  {Kluyver}}, \bibinfo {author} {\bibfnamefont {B.}~\bibnamefont
  {Ragan-Kelley}}, \bibinfo {author} {\bibfnamefont {F.}~\bibnamefont
  {P{\'e}rez}}, \bibinfo {author} {\bibfnamefont {B.}~\bibnamefont {Granger}},
  \bibinfo {author} {\bibfnamefont {M.}~\bibnamefont {Bussonnier}}, \bibinfo
  {author} {\bibfnamefont {J.}~\bibnamefont {Frederic}}, \bibinfo {author}
  {\bibfnamefont {K.}~\bibnamefont {Kelley}}, \bibinfo {author} {\bibfnamefont
  {J.}~\bibnamefont {Hamrick}}, \bibinfo {author} {\bibfnamefont
  {J.}~\bibnamefont {Grout}}, \bibinfo {author} {\bibfnamefont
  {S.}~\bibnamefont {Corlay}},  \emph {et~al.},\ }in\ \href@noop {} {\emph
  {\bibinfo {booktitle} {Positioning and power in academic publishing: Players,
  agents and agendas}}}\ (\bibinfo  {publisher} {IOS press},\ \bibinfo {year}
  {2016})\ pp.\ \bibinfo {pages} {87--90}\BibitemShut {NoStop}%
\bibitem [{\citenamefont {Kolevatov}\ \emph {et~al.}(2019)\citenamefont
  {Kolevatov}, \citenamefont {Schanzer},\ and\ \citenamefont
  {B{\"o}ni}}]{kolevatov2019neutron}%
  \BibitemOpen
  \bibfield  {author} {\bibinfo {author} {\bibfnamefont {R.}~\bibnamefont
  {Kolevatov}}, \bibinfo {author} {\bibfnamefont {C.}~\bibnamefont {Schanzer}},
  \ and\ \bibinfo {author} {\bibfnamefont {P.}~\bibnamefont {B{\"o}ni}},\
  }\href@noop {} {\bibfield  {journal} {\bibinfo  {journal} {Nuclear
  Instruments and Methods in Physics Research Section A: Accelerators,
  Spectrometers, Detectors and Associated Equipment}\ }\textbf {\bibinfo
  {volume} {922}},\ \bibinfo {pages} {98} (\bibinfo {year} {2019})}\BibitemShut
  {NoStop}%
\bibitem [{\citenamefont {Schultz}\ \emph {et~al.}(2005)\citenamefont
  {Schultz}, \citenamefont {Thiyagarajan}, \citenamefont {Hodges},
  \citenamefont {Rehm}, \citenamefont {Myles}, \citenamefont {Langan},\ and\
  \citenamefont {Mesecar}}]{schultz2005conceptual}%
  \BibitemOpen
  \bibfield  {author} {\bibinfo {author} {\bibfnamefont {A.~J.}\ \bibnamefont
  {Schultz}}, \bibinfo {author} {\bibfnamefont {P.}~\bibnamefont
  {Thiyagarajan}}, \bibinfo {author} {\bibfnamefont {J.~P.}\ \bibnamefont
  {Hodges}}, \bibinfo {author} {\bibfnamefont {C.}~\bibnamefont {Rehm}},
  \bibinfo {author} {\bibfnamefont {D.~A.}\ \bibnamefont {Myles}}, \bibinfo
  {author} {\bibfnamefont {P.}~\bibnamefont {Langan}}, \ and\ \bibinfo {author}
  {\bibfnamefont {A.~D.}\ \bibnamefont {Mesecar}},\ }\href@noop {} {\bibfield
  {journal} {\bibinfo  {journal} {Journal of applied crystallography}\ }\textbf
  {\bibinfo {volume} {38}},\ \bibinfo {pages} {964} (\bibinfo {year}
  {2005})}\BibitemShut {NoStop}%
\bibitem [{\citenamefont {Batz}\ \emph {et~al.}(2005)\citenamefont {Batz},
  \citenamefont {Bae{\ss}ler}, \citenamefont {Heil}, \citenamefont {Otten},
  \citenamefont {Rudersdorf}, \citenamefont {Schmiedeskamp}, \citenamefont
  {Sobolev},\ and\ \citenamefont {Wolf}}]{batz20053he}%
  \BibitemOpen
  \bibfield  {author} {\bibinfo {author} {\bibfnamefont {M.}~\bibnamefont
  {Batz}}, \bibinfo {author} {\bibfnamefont {S.}~\bibnamefont {Bae{\ss}ler}},
  \bibinfo {author} {\bibfnamefont {W.}~\bibnamefont {Heil}}, \bibinfo {author}
  {\bibfnamefont {E.}~\bibnamefont {Otten}}, \bibinfo {author} {\bibfnamefont
  {D.}~\bibnamefont {Rudersdorf}}, \bibinfo {author} {\bibfnamefont
  {J.}~\bibnamefont {Schmiedeskamp}}, \bibinfo {author} {\bibfnamefont
  {Y.}~\bibnamefont {Sobolev}}, \ and\ \bibinfo {author} {\bibfnamefont
  {M.}~\bibnamefont {Wolf}},\ }\href@noop {} {\bibfield  {journal} {\bibinfo
  {journal} {Journal of Research of the National Institute of Standards and
  Technology}\ }\textbf {\bibinfo {volume} {110}},\ \bibinfo {pages} {293}
  (\bibinfo {year} {2005})}\BibitemShut {NoStop}%
\bibitem [{\citenamefont {Stahn}\ and\ \citenamefont
  {Glavic}(2017)}]{stahn2017efficient}%
  \BibitemOpen
  \bibfield  {author} {\bibinfo {author} {\bibfnamefont {J.}~\bibnamefont
  {Stahn}}\ and\ \bibinfo {author} {\bibfnamefont {A.}~\bibnamefont {Glavic}},\
  }in\ \href@noop {} {\emph {\bibinfo {booktitle} {Journal of Physics:
  Conference Series}}},\ Vol.\ \bibinfo {volume} {862}\ (\bibinfo
  {organization} {IOP Publishing},\ \bibinfo {year} {2017})\ p.\ \bibinfo
  {pages} {012007}\BibitemShut {NoStop}%
\bibitem [{\citenamefont {DiJulio}\ \emph {et~al.}(2022)\citenamefont
  {DiJulio}, \citenamefont {Santoro}, \citenamefont {Devishvili}, \citenamefont
  {Khaplanov}, \citenamefont {Kolevatov}, \citenamefont {Mag{\'a}n},
  \citenamefont {Miller},\ and\ \citenamefont
  {Muhrer}}]{dijulio2022measurements}%
  \BibitemOpen
  \bibfield  {author} {\bibinfo {author} {\bibfnamefont {D.}~\bibnamefont
  {DiJulio}}, \bibinfo {author} {\bibfnamefont {V.}~\bibnamefont {Santoro}},
  \bibinfo {author} {\bibfnamefont {A.}~\bibnamefont {Devishvili}}, \bibinfo
  {author} {\bibfnamefont {A.}~\bibnamefont {Khaplanov}}, \bibinfo {author}
  {\bibfnamefont {R.}~\bibnamefont {Kolevatov}}, \bibinfo {author}
  {\bibfnamefont {M.}~\bibnamefont {Mag{\'a}n}}, \bibinfo {author}
  {\bibfnamefont {T.}~\bibnamefont {Miller}}, \ and\ \bibinfo {author}
  {\bibfnamefont {G.}~\bibnamefont {Muhrer}},\ }\href@noop {} {\bibfield
  {journal} {\bibinfo  {journal} {Nuclear Instruments and Methods in Physics
  Research Section A: Accelerators, Spectrometers, Detectors and Associated
  Equipment}\ }\textbf {\bibinfo {volume} {1025}},\ \bibinfo {pages} {166088}
  (\bibinfo {year} {2022})}\BibitemShut {NoStop}%
\bibitem [{\citenamefont {Gentile}\ \emph {et~al.}(2017)\citenamefont
  {Gentile}, \citenamefont {Nacher}, \citenamefont {Saam},\ and\ \citenamefont
  {Walker}}]{gentile2017optically}%
  \BibitemOpen
  \bibfield  {author} {\bibinfo {author} {\bibfnamefont {T.~R.}\ \bibnamefont
  {Gentile}}, \bibinfo {author} {\bibfnamefont {P.}~\bibnamefont {Nacher}},
  \bibinfo {author} {\bibfnamefont {B.}~\bibnamefont {Saam}}, \ and\ \bibinfo
  {author} {\bibfnamefont {T.}~\bibnamefont {Walker}},\ }\href@noop {}
  {\bibfield  {journal} {\bibinfo  {journal} {Reviews of modern physics}\
  }\textbf {\bibinfo {volume} {89}},\ \bibinfo {pages} {045004} (\bibinfo
  {year} {2017})}\BibitemShut {NoStop}%
\bibitem [{\citenamefont {Ye}\ \emph {et~al.}(2013)\citenamefont {Ye},
  \citenamefont {Gentile}, \citenamefont {Anderson}, \citenamefont {Broholm},
  \citenamefont {Chen}, \citenamefont {DeLand}, \citenamefont {Erwin},
  \citenamefont {Fu}, \citenamefont {Fuller}, \citenamefont {Kirchhoff} \emph
  {et~al.}}]{ye2013wide}%
  \BibitemOpen
  \bibfield  {author} {\bibinfo {author} {\bibfnamefont {Q.}~\bibnamefont
  {Ye}}, \bibinfo {author} {\bibfnamefont {T.~R.}\ \bibnamefont {Gentile}},
  \bibinfo {author} {\bibfnamefont {J.}~\bibnamefont {Anderson}}, \bibinfo
  {author} {\bibfnamefont {C.}~\bibnamefont {Broholm}}, \bibinfo {author}
  {\bibfnamefont {W.}~\bibnamefont {Chen}}, \bibinfo {author} {\bibfnamefont
  {Z.}~\bibnamefont {DeLand}}, \bibinfo {author} {\bibfnamefont {R.~W.}\
  \bibnamefont {Erwin}}, \bibinfo {author} {\bibfnamefont {C.}~\bibnamefont
  {Fu}}, \bibinfo {author} {\bibfnamefont {J.}~\bibnamefont {Fuller}}, \bibinfo
  {author} {\bibfnamefont {A.}~\bibnamefont {Kirchhoff}},  \emph {et~al.},\
  }\href@noop {} {\bibfield  {journal} {\bibinfo  {journal} {Physics Procedia}\
  }\textbf {\bibinfo {volume} {42}},\ \bibinfo {pages} {206} (\bibinfo {year}
  {2013})}\BibitemShut {NoStop}%
\bibitem [{\citenamefont {Chen}\ \emph {et~al.}(2014)\citenamefont {Chen},
  \citenamefont {Gentile}, \citenamefont {Ye}, \citenamefont {Walker},\ and\
  \citenamefont {Babcock}}]{chen2014limits}%
  \BibitemOpen
  \bibfield  {author} {\bibinfo {author} {\bibfnamefont {W.}~\bibnamefont
  {Chen}}, \bibinfo {author} {\bibfnamefont {T.}~\bibnamefont {Gentile}},
  \bibinfo {author} {\bibfnamefont {Q.}~\bibnamefont {Ye}}, \bibinfo {author}
  {\bibfnamefont {T.}~\bibnamefont {Walker}}, \ and\ \bibinfo {author}
  {\bibfnamefont {E.}~\bibnamefont {Babcock}},\ }\href@noop {} {\bibfield
  {journal} {\bibinfo  {journal} {Journal of applied physics}\ }\textbf
  {\bibinfo {volume} {116}},\ \bibinfo {pages} {014903} (\bibinfo {year}
  {2014})}\BibitemShut {NoStop}%
\bibitem [{\citenamefont {Salhi}\ \emph {et~al.}(2019)\citenamefont {Salhi},
  \citenamefont {Babcock}, \citenamefont {Bing{\"o}l}, \citenamefont
  {Bussmann}, \citenamefont {Kammerling}, \citenamefont {Ossovyi},
  \citenamefont {Heynen}, \citenamefont {Deng}, \citenamefont {Hutanu},
  \citenamefont {Masalovich} \emph {et~al.}}]{salhi2019situ}%
  \BibitemOpen
  \bibfield  {author} {\bibinfo {author} {\bibfnamefont {Z.}~\bibnamefont
  {Salhi}}, \bibinfo {author} {\bibfnamefont {E.}~\bibnamefont {Babcock}},
  \bibinfo {author} {\bibfnamefont {K.}~\bibnamefont {Bing{\"o}l}}, \bibinfo
  {author} {\bibfnamefont {K.}~\bibnamefont {Bussmann}}, \bibinfo {author}
  {\bibfnamefont {H.}~\bibnamefont {Kammerling}}, \bibinfo {author}
  {\bibfnamefont {V.}~\bibnamefont {Ossovyi}}, \bibinfo {author} {\bibfnamefont
  {A.}~\bibnamefont {Heynen}}, \bibinfo {author} {\bibfnamefont
  {H.}~\bibnamefont {Deng}}, \bibinfo {author} {\bibfnamefont {V.}~\bibnamefont
  {Hutanu}}, \bibinfo {author} {\bibfnamefont {S.}~\bibnamefont {Masalovich}},
  \emph {et~al.},\ }in\ \href@noop {} {\emph {\bibinfo {booktitle} {Journal of
  Physics: Conference Series}}},\ Vol.\ \bibinfo {volume} {1316}\ (\bibinfo
  {organization} {IOP Publishing},\ \bibinfo {year} {2019})\ p.\ \bibinfo
  {pages} {012009}\BibitemShut {NoStop}%
\bibitem [{\citenamefont {Buras}\ and\ \citenamefont
  {Gerward}(1975)}]{buras1975relations}%
  \BibitemOpen
  \bibfield  {author} {\bibinfo {author} {\bibfnamefont {B.}~\bibnamefont
  {Buras}}\ and\ \bibinfo {author} {\bibfnamefont {L.}~\bibnamefont
  {Gerward}},\ }\href@noop {} {\bibfield  {journal} {\bibinfo  {journal} {Acta
  Crystallographica Section A: Crystal Physics, Diffraction, Theoretical and
  General Crystallography}\ }\textbf {\bibinfo {volume} {31}},\ \bibinfo
  {pages} {372} (\bibinfo {year} {1975})}\BibitemShut {NoStop}%
\bibitem [{\citenamefont {Zhao}\ \emph {et~al.}(2013)\citenamefont {Zhao},
  \citenamefont {Robertson}, \citenamefont {Herwig}, \citenamefont
  {Gallmeier},\ and\ \citenamefont {Riemer}}]{zhao2013optimizing}%
  \BibitemOpen
  \bibfield  {author} {\bibinfo {author} {\bibfnamefont {J.}~\bibnamefont
  {Zhao}}, \bibinfo {author} {\bibfnamefont {J.}~\bibnamefont {Robertson}},
  \bibinfo {author} {\bibfnamefont {K.~W.}\ \bibnamefont {Herwig}}, \bibinfo
  {author} {\bibfnamefont {F.~X.}\ \bibnamefont {Gallmeier}}, \ and\ \bibinfo
  {author} {\bibfnamefont {B.~W.}\ \bibnamefont {Riemer}},\ }\href@noop {}
  {\bibfield  {journal} {\bibinfo  {journal} {Review of Scientific
  Instruments}\ }\textbf {\bibinfo {volume} {84}},\ \bibinfo {pages} {125104}
  (\bibinfo {year} {2013})}\BibitemShut {NoStop}%
\bibitem [{\citenamefont {Batkov}\ \emph {et~al.}(2013)\citenamefont {Batkov},
  \citenamefont {Takibayev}, \citenamefont {Zanini},\ and\ \citenamefont
  {Mezei}}]{batkov2013unperturbed}%
  \BibitemOpen
  \bibfield  {author} {\bibinfo {author} {\bibfnamefont {K.}~\bibnamefont
  {Batkov}}, \bibinfo {author} {\bibfnamefont {A.}~\bibnamefont {Takibayev}},
  \bibinfo {author} {\bibfnamefont {L.}~\bibnamefont {Zanini}}, \ and\ \bibinfo
  {author} {\bibfnamefont {F.}~\bibnamefont {Mezei}},\ }\href@noop {}
  {\bibfield  {journal} {\bibinfo  {journal} {Nuclear Instruments and Methods
  in Physics Research Section A: Accelerators, Spectrometers, Detectors and
  Associated Equipment}\ }\textbf {\bibinfo {volume} {729}},\ \bibinfo {pages}
  {500} (\bibinfo {year} {2013})}\BibitemShut {NoStop}%
\bibitem [{\citenamefont {Rowe}(2005)}]{rowe2005analysis}%
  \BibitemOpen
  \bibfield  {author} {\bibinfo {author} {\bibfnamefont {J.~M.}\ \bibnamefont
  {Rowe}},\ }in\ \href@noop {} {\emph {\bibinfo {booktitle} {Joint Mtg.
  National Organization of Test, Research, and Training Reactors and the
  International Group on Research Reactors}}}\ (\bibinfo {year}
  {2005})\BibitemShut {NoStop}%
\bibitem [{\citenamefont {Andersen}\ \emph {et~al.}(2018)\citenamefont
  {Andersen}, \citenamefont {Bertelsen}, \citenamefont {Zanini}, \citenamefont
  {Klinkby}, \citenamefont {Sch{\"o}nfeldt}, \citenamefont {Bentley},\ and\
  \citenamefont {Saroun}}]{andersen2018optimization}%
  \BibitemOpen
  \bibfield  {author} {\bibinfo {author} {\bibfnamefont {K.~H.}\ \bibnamefont
  {Andersen}}, \bibinfo {author} {\bibfnamefont {M.}~\bibnamefont {Bertelsen}},
  \bibinfo {author} {\bibfnamefont {L.}~\bibnamefont {Zanini}}, \bibinfo
  {author} {\bibfnamefont {E.~B.}\ \bibnamefont {Klinkby}}, \bibinfo {author}
  {\bibfnamefont {T.}~\bibnamefont {Sch{\"o}nfeldt}}, \bibinfo {author}
  {\bibfnamefont {P.~M.}\ \bibnamefont {Bentley}}, \ and\ \bibinfo {author}
  {\bibfnamefont {J.}~\bibnamefont {Saroun}},\ }\href@noop {} {\bibfield
  {journal} {\bibinfo  {journal} {Journal of applied crystallography}\ }\textbf
  {\bibinfo {volume} {51}},\ \bibinfo {pages} {264} (\bibinfo {year}
  {2018})}\BibitemShut {NoStop}%
\bibitem [{swi()}]{swissneutronics}%
  \BibitemOpen
  \href@noop {} {}\bibinfo {howpublished}
  {\url{https://www.swissneutronics.ch/products/neutron-supermirrors/}},\
  \bibinfo {note} {accessed: 2024-07-19}\BibitemShut {NoStop}%
\bibitem [{\citenamefont {Jacobsen}\ \emph {et~al.}(2013)\citenamefont
  {Jacobsen}, \citenamefont {Lieutenant}, \citenamefont {Zendler},\ and\
  \citenamefont {Lefmann}}]{jacobsen2013bi}%
  \BibitemOpen
  \bibfield  {author} {\bibinfo {author} {\bibfnamefont {H.}~\bibnamefont
  {Jacobsen}}, \bibinfo {author} {\bibfnamefont {K.}~\bibnamefont
  {Lieutenant}}, \bibinfo {author} {\bibfnamefont {C.}~\bibnamefont {Zendler}},
  \ and\ \bibinfo {author} {\bibfnamefont {K.}~\bibnamefont {Lefmann}},\
  }\href@noop {} {\bibfield  {journal} {\bibinfo  {journal} {Nuclear
  Instruments and Methods in Physics Research Section A: Accelerators,
  Spectrometers, Detectors and Associated Equipment}\ }\textbf {\bibinfo
  {volume} {717}},\ \bibinfo {pages} {69} (\bibinfo {year} {2013})}\BibitemShut
  {NoStop}%
\end{thebibliography}
\end{document}